\documentclass[a4paper,11pt]{article}
\usepackage{jcappub} 
\usepackage{lineno}
\usepackage{siunitx}
\usepackage[normalem]{ulem}
\usepackage{orcidlink}

\newcommand{\gthree}{I\hspace{-.1em}I\hspace{-.1em}I }
\newcommand{\edit}[1]{{#1}}

\title{\boldmath Development and characterization of a millimeter-wave circular polarization observation method using a quarter-wave plate with Simons Array}

\usepackage{orcidlink}

\usepackage{orcidlink}

\author[a]{Kana Sakaguri\,\orcidlink{0000-0001-5667-8118},}
\author[b]{Darcy R. Barron\,\orcidlink{0000-0002-1623-5651},}
\author[c,d]{Yuji Chinone\,\orcidlink{0000-0002-3266-857X},}
\author[e]{Matthew Ryan Chu\,\orcidlink{0000-0001-5884-911X},}
\author[f]{Nicole Farias\,\orcidlink{0000-0002-4101-2513},}
\author[c]{Takuro Fujino\,\orcidlink{0000-0002-1211-7850},}
\author[g,c]{Masaya Hasegawa\,\orcidlink{0000-0003-1443-1082},}
\author[p]{Masashi Hazumi\,\orcidlink{0000-0001-6830-8309},}
\author[g]{Haruaki Hirose\,\orcidlink{0000-0003-3336-134X},}
\author[h]{Jennifer Ito\,\orcidlink{0009-0008-3216-4022},}
\author[i,j]{Oliver Jeong\,\orcidlink{0000-0001-5893-7697},}
\author[k]{Daisuke Kaneko\,\orcidlink{0000-0003-3917-086X},}
\author[e]{Brian Keating,}
\author[a,l,d,m]{Akito Kusaka\,\orcidlink{0009-0004-9631-2451},}
\author[f,m]{Adrian T. Lee\,\orcidlink{0000-0003-3106-3218},}
\author[f]{Lindsay Ng Lowry,}
\author[n]{Richard Plambeck\,\orcidlink{0000-0001-6765-9609},}
\author[e]{Shahed Shayan Arani\,\orcidlink{0009-0005-4168-2858},}
\author[o]{Praween Siritanasak\,\orcidlink{0000-0001-6830-1537},}
\author[a]{and Satoru Takakura\,\orcidlink{0000-0001-9461-7519}}

\affiliation[a]{Department of Physics, The University of Tokyo,\\
Tokyo 113-0033, Japan}

\affiliation[b]{Department of Physics and Astronomy, University of New Mexico,\\
Albuquerque, NM 87131, USA}

\affiliation[c]{International Center for Quantum-field Measurement Systems
for Studies of the Universe and Particles (QUP),
High Energy Accelerator Research Organization (KEK),\\
Tsukuba, Ibaraki 305-0801, Japan}

\affiliation[d]{Kavli Institute for the Physics and Mathematics of the Universe
(WPI), UTIAS, The University of Tokyo,\\
Kashiwa, Chiba 277-8583, Japan}

\affiliation[e]{Department of Physics, University of California, San Diego,\\
La Jolla, CA 92093, USA}

\affiliation[f]{Department of Physics, University of California, Berkeley,\\
Berkeley, CA 94720, USA}

\affiliation[g]{Institute of Particle and Nuclear Studies (IPNS),
High Energy Accelerator Research Organization (KEK),\\
Tsukuba, Ibaraki 305-0801, Japan}

\affiliation[h]{Department of Physics, Westmont College,\\
Santa Barbara, CA 93108, USA}

\affiliation[i]{CNRS-UCB International Research Laboratory,
Centre Pierre Bin\'etruy, IRL 2007, CPB-IN2P3,\\
Berkeley, CA 94720, USA}

\affiliation[j]{Lawrence Berkeley National Laboratory,\\
1 Cyclotron Road, Berkeley, CA 94720, USA}

\affiliation[k]{Accelerator Laboratory,
High Energy Accelerator Research Organization (KEK),\\
Tsukuba, Ibaraki 305-0801, Japan}

\affiliation[l]{Research Center for the Early Universe,
School of Science, The University of Tokyo,\\
Tokyo 113-0033, Japan}

\affiliation[m]{Physics Division, Lawrence Berkeley National Laboratory,\\
Berkeley, CA 94720, USA}

\affiliation[n]{Radio Astronomy Laboratory,
University of California, Berkeley,\\
Berkeley, CA 94720, USA}

\affiliation[o]{National Astronomical Research Institute of Thailand,\\
Chiang Mai 50180, Thailand}

\affiliation[p]{Department of Physics and Center for High Energy and High Field Physics (CHiP),
National Central University,\\
Taoyuan City, Taiwan}

\emailAdd{kana.sakaguri@yale.edu}

\abstract{\edit{We have developed and demonstrated a method for measuring circular polarization at \SI{90}{GHz} by combining a quarter-wave plate (QWP) and a continuously rotating half-wave plate (HWP).} While many bolometric experiments at millimeter wavelengths focus on measuring linear polarization, observations of circular polarization in millimeter waves may open a new window into astrophysics and cosmology.
We present a novel approach to circular-polarization search using linear-polarization-sensitive detectors by incorporating a QWP in the optics. The Simons Array was originally designed for precise linear polarization measurements, and we subsequently installed a QWP \edit{in front of the receiver} targeting \SI{90}{GHz}. The QWP converts incident circular polarization to linear polarization, allowing detection with our detectors.
\edit{Laboratory measurements demonstrate a QWP conversion efficiency of 99.0\%.}
We apply our method to observations of unpolarized and linearly polarized bright sources, Jupiter and Tau~A.
\edit{Using these observations, we establish the calibration and characterize the instrumental leakages relevant to circular polarization measurements. The measured polarization efficiency adopted for Stokes $V$ is $46\pm2$\%, and the measured U-to-V leakage is $-11.7\pm5.1$\%. We measure a circular polarization fraction of $V/I=0.78\pm0.42$\% for Tau~A, showing no significant detection and providing a consistency check of the calibration and systematic-uncertainty treatment. These results establish the instrument response model and calibration parameters used for the companion circular-polarization source search.}
}

\begin{document}
\maketitle
\flushbottom

\section{Introduction}
\label{sec:intro}
Observations of millimeter-wave polarization provide rich information for understanding the universe. One of the main observation targets at millimeter waves is the cosmic microwave background (CMB) linear polarization. In particular, CMB $B$-modes generated by primordial gravitational waves are expected to be a powerful probe of cosmic inflation at the beginning of the universe~\cite{PhysRevLett.78.2054,PhysRevLett.78.2058}.

The CMB community is growing, including both ground- and space-based experiments.
Current CMB observations mainly aim to push the constraint of the tensor-to-scalar ratio $r$ further or detect the $B$-mode polarization induced by primordial gravitational waves. 
These CMB experiments utilize the technology of an incoherent detector, which directly measures the intensity of incident CMB signals. Thanks to their high sensitivity and low noise, Transition-Edge Sensor (TES) bolometers are often used~\cite{Duff2016,Stevens2020}. By increasing the number of bolometers deployed, these telescopes aim to achieve high sensitivity.
For example, the Simons Array (SA) experiment, which is the successor of the \textsc{Polarbear} experiment~\cite{PB_2012,Kermish2012phd,Adachi_2020,Adachi_2020E} at the Atacama Desert in Chile, was designed to measure the CMB linear polarization with high accuracy~\cite{SA_design}.

Though the mainstream of recent millimeter-wave polarization experiments focuses on linear polarization, observations of CMB circular polarization can probe a variety of effects~\cite{eimer2023class,R.Mainini_2013_MIPOL,Nagy_2017_SPIDER,Padilla_2020_CLASS}, such as galactic synchrotron emission~\cite{Planck_2020e,Legg_1968CP,King2016_CPsync,eimer2023class}, Faraday conversion by Population (Pop)~\gthree stars~\cite{De_2015Pop3}, scattering with the cosmic neutrino background~\cite{Mohammadi2014}, photon-photon forward scattering~\cite{Hoseinpour2020}, and Lorentz violation~\cite{Caloni_2023}. The search for circular polarization in astronomical sources may reveal magnetic field structures~\cite{Faradayconv2011,wiesemeyer2011,Ruszkowski_2002}.
In addition, the Earth's atmospheric emission is circularly polarized due to the Zeeman effect~\cite{Lenoir1968,Keating1998,HANANY2003,Spinelli2011}.
The CLASS and \textsc{Polarbear} experiments detected this circular polarization~\cite{Petroff_2020,TakuroCP}. CLASS uses a variable-delay polarization modulator, whereas \textsc{Polarbear} takes advantage of the nonidealities of a continuously rotating half-wave plate (HWP)~\cite{Chuss:12,Harrington2018,fujinohwp_23,Nagy_2017_SPIDER}.

Among several methods to measure circular polarization in millimeter waves~\cite{Petroff_2020,Nagy_2017_SPIDER,eimer2023class,TakuroCP}, our approach is to use a quarter-wave plate (QWP)~\cite{vanVliet1981}.
The QWP gives a phase difference of $\lambda/4$ between the two linear components of the input signals. \edit{It converts circular polarization into linear polarization, allowing Stokes $V$ to be modulated by the HWP and measured with the existing receiver.} We fabricated the QWP made of Rexolite~1422, cross-linked polystyrene.\footnote{C-Lec Plastics, Inc., Philadelphia, PA, USA.} We cut grooves into both sides of the plate, adjusting the thickness to optimize the QWP for a \SI{90}{GHz} circular polarization search.

The QWP was installed in the SA telescope that hosts the \textsc{Polarbear}-2b~(PB-2b) receiver. PB-2b is the second receiver in the SA, which was designed to measure the linear polarization of the CMB~\cite{SA_design}. It did not originally have circular polarization as a science goal. It contains 7588 TES bolometers 
and utilizes a cryogenic continuously rotating HWP for modulation of the incoming linear polarization signal~\cite{Hill2020hwp}.
The QWP is located on the sky side of the receiver and operates at ambient air temperature. \edit{By converting Stokes $V$ before HWP modulation, the QWP enables the existing system to measure circular polarization without a drastic change to the telescope structure or analysis pipeline.}


\edit{The main goal of this analysis is to characterize the instrumental response and determine the calibration parameters relevant to circular polarization measurements, including those associated with nonidealities of the QWP and other instrumental components.}
Tau~A, also known as the Crab Nebula, and Jupiter observations are used for characterization and calibration. They have high-intensity signals, and Tau~A has large linear polarization of about 8.8\% at our target frequency~\cite{TauA_linear_2010}.
Since the QWP converts the source circular polarization to linear polarization, the overall analysis pipeline is similar to those used for linear polarization analysis, such as that used in the analysis of \textsc{Polarbear}-2a (PB-2a) observations of Tau~A~\cite{2aTauA}.
We developed the analysis methodology and the calibration method by utilizing the QWP, and we identified systematic errors unique to this type of measurement. We estimated the efficiency and leakages from Tau~A and Jupiter observations.
The analysis method described here is used for searching for circular polarization in astronomical sources, for which we apply blind analysis. The results are presented in a companion paper~\cite{mypaper} (hereafter Paper II).

This paper presents the development of the QWP and characterizes circular polarization search with the SA.
Section~\ref{sec:QWP} describes the method for the circular polarization search with the QWP and its evaluation at a lab. Section~\ref{subsec:dataset} presents the Tau~A and Jupiter datasets. In section~\ref{sec:analysis}, we report the analysis from time-ordered data processing to map making. We discuss the results of on-sky characterization with Tau~A and Jupiter observations in section~\ref{subsec:characterization}. We conclude the paper in section~\ref{sec:conclusion}.

\section{Method}
\label{sec:QWP}

\subsection{Overview of the instrumentation for circular polarization measurements}
\label{subsec:method overview}
\begin{figure}[tbp]
\centering
\includegraphics[width=.45\textwidth]{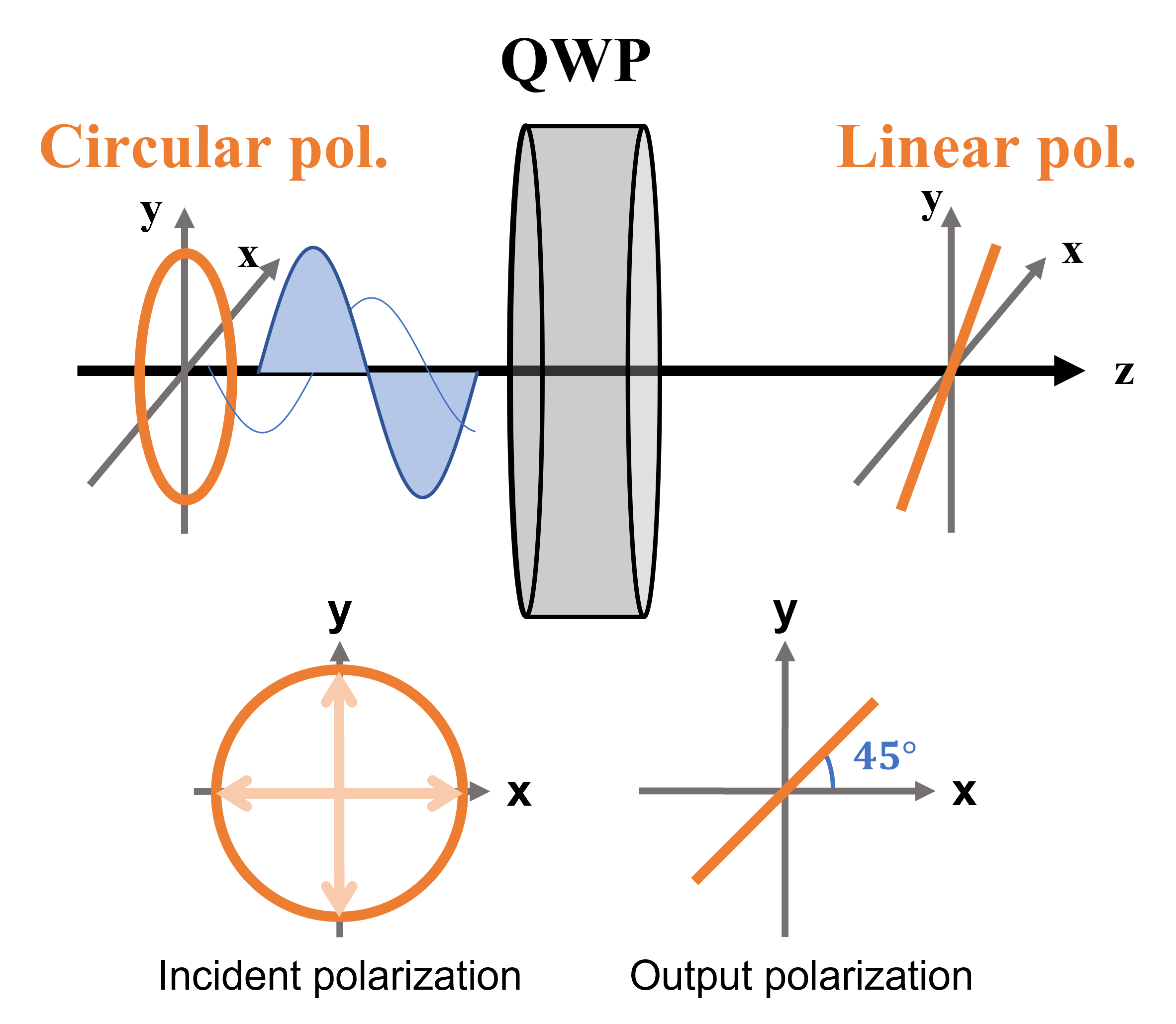}
\qquad
\includegraphics[width=.45\textwidth]{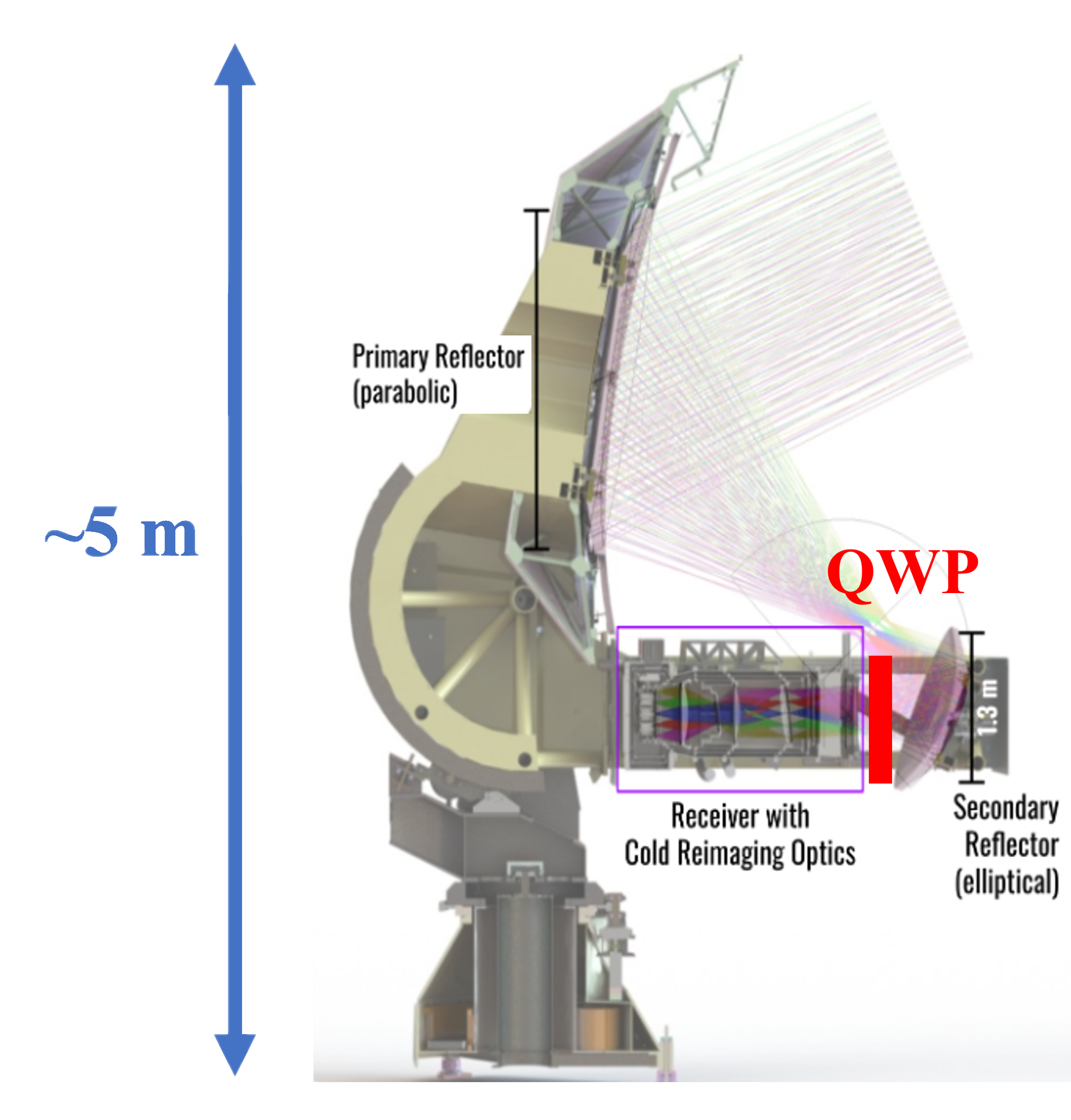}
\caption{Left: Schematics of the QWP. The circularly polarized light propagates to the z-axis
direction. The QWP optical axis is aligned to the x (y)-axis. Incident circularly polarized light is decomposed into two orthogonal linear polarization components aligned with the axes. The QWP introduces a 90-degree phase delay between the two components, converting circular polarization into linear polarization at the output. Right: Cross section of a 3D CAD model of the telescope. Figure based on \cite{Lowry2021}.\label{fig:QWP overview schematics}}
\end{figure}

The QWP is the main optical element for the PB-2b circular polarization
search.
\edit{When aligned with the reference axes, the QWP converts input Stokes $V$ into Stokes $U$} (figure~\ref{fig:QWP overview schematics} left). Details about our QWP are described in section~\ref{subsec:mechanism}.

We install an ambient temperature QWP on the sky side of the
PB-2b receiver. Figure~\ref{fig:QWP overview schematics} right shows the cross section view of the telescope. \edit{Because the PB-2b receiver and its modulation scheme are designed to measure linear polarization, the QWP enables the measurement of sky circular polarization by converting Stokes $V$ into linear polarization.}

\subsection{Simons Array instrument}
\label{subsec:instrument}
The SA is a ground-based experiment on the Atacama plateau in Chile that aims to measure \edit{polarization of millimeter-wave signals}. PB-2a is the first receiver in the SA, while PB-2b is the second~\cite{SA_design,Praween2018,Lowry2021,Jen2022}. The two receivers are very similar in their designs. The QWP discussed here is used with PB-2b.

\begin{figure}[tbp]
    \centering
    \includegraphics[width=0.63\linewidth, keepaspectratio]{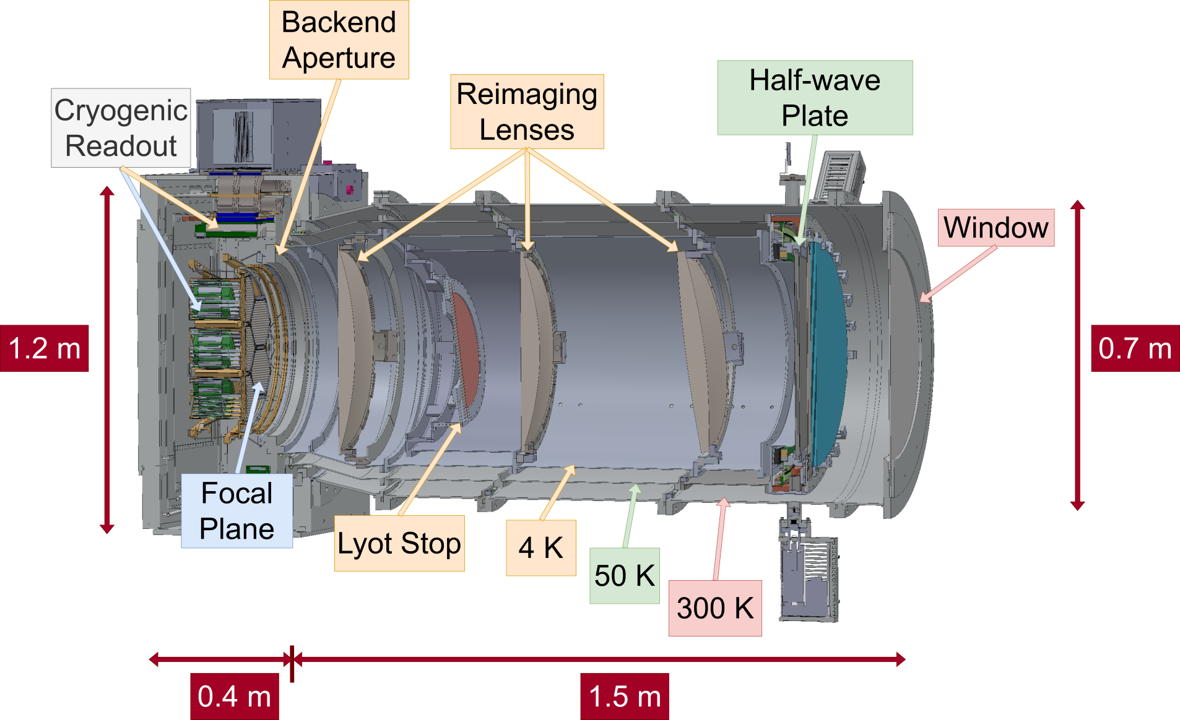}
    \caption{Cross section of the PB-2b receiver, with key elements labeled~\cite{Jen2022}.}
    \label{fig:receiver and QWP}
\end{figure}

The telescope employs an off-axis Gregorian optical design. The design of the reflective optics is identical to that of the \textsc{Polarbear}-1 (PB-1) telescope, the predecessor to SA~\cite{PB_2012}. \edit{A cold Lyot stop in the receiver optics defines an effective primary aperture of \SI{2.5}{m}, projected along boresight onto the central, high-precision panel of the primary mirror.} It satisfies the Mizuguchi-Dragone condition, in order to mitigate cross-polarization and astigmatism effects~\cite{TRAN20031091}.
The field of view is circular with 4.5-degree diameter~\cite{Matsuda2017,Lowry2021}, and an angular resolution is designed to be 5.2 (3.5) arc minutes full width at half maximum at 90 (150)\,GHz~\cite{Tomaru2012}.
The receiver is cooled by sorption refrigerators~\cite{BHATIA2000685}. Figure~\ref{fig:receiver and QWP} shows the PB-2b receiver. It consists mainly of a \SI{0.5}{m} diameter vacuum window, three lenses cooled to \SI{4}{\kelvin}, and a \SI{0.3}{\kelvin} focal plane. A cryogenic HWP is located between the vacuum window and the lenses~\cite{Hill2018hwp}. The QWP is installed \SI{150}{mm} away from the window at ambient temperature. It is designed to cover all the light rays without intercepting the optical path between the primary and secondary mirrors.


The PB-2b design is almost the same as that of PB-2a~\cite{2aTauA}. PB-2b, however, uses a cryogenic HWP that consists of three-stacked sapphire with two-layer anti-reflection (AR) coatings~\cite{Hill2020hwp,Sakaguri:24,11474334}.
The HWP rotates at \SI{2}{Hz} and modulates linear polarization signals to mitigate atmospheric 1/f noise. It prevents systematics arising from different responses in pairs of polarimeters by rotating continuously~\cite{Takakura_2017,Kusakahwp,Essinger16}.
The PB-2b HWP is operated at a cryogenic temperature of \SI{50}{K} to minimize the effects of thermal radiation. It is located near the telescope's Gregorian focus. Its drive system consists of a superconducting magnetic bearing and a synchronous electromagnetic motor~\cite{Hill2020hwp}.

The focal plane consists of dichroic detectors, with each spectral band (centered at 90 and \SI{150}{GHz}) split into orthogonal linear polarizations. Each observing band has a fractional bandwidth of approximately
30\%. The receiver is equipped with \edit{7588 bolometers}. These bolometers are on seven wafers on the focal plane, each containing 271 pixels.
Each pixel features a lenslet-coupled sinuous antenna for orthogonal polarizations, which feeds two superconducting TES bolometers per frequency band.
These TES bolometers have a transition temperature of about \SI{450}{mK}~\cite{Westbrook2018}.

\subsection{Quarter-wave plate}
\label{subsec:mechanism}
\subsubsection{Design and fabrication}
\label{subsubsec:design and fab}
\begin{figure}[tbp]
\centering
\includegraphics[width=.35\textwidth]{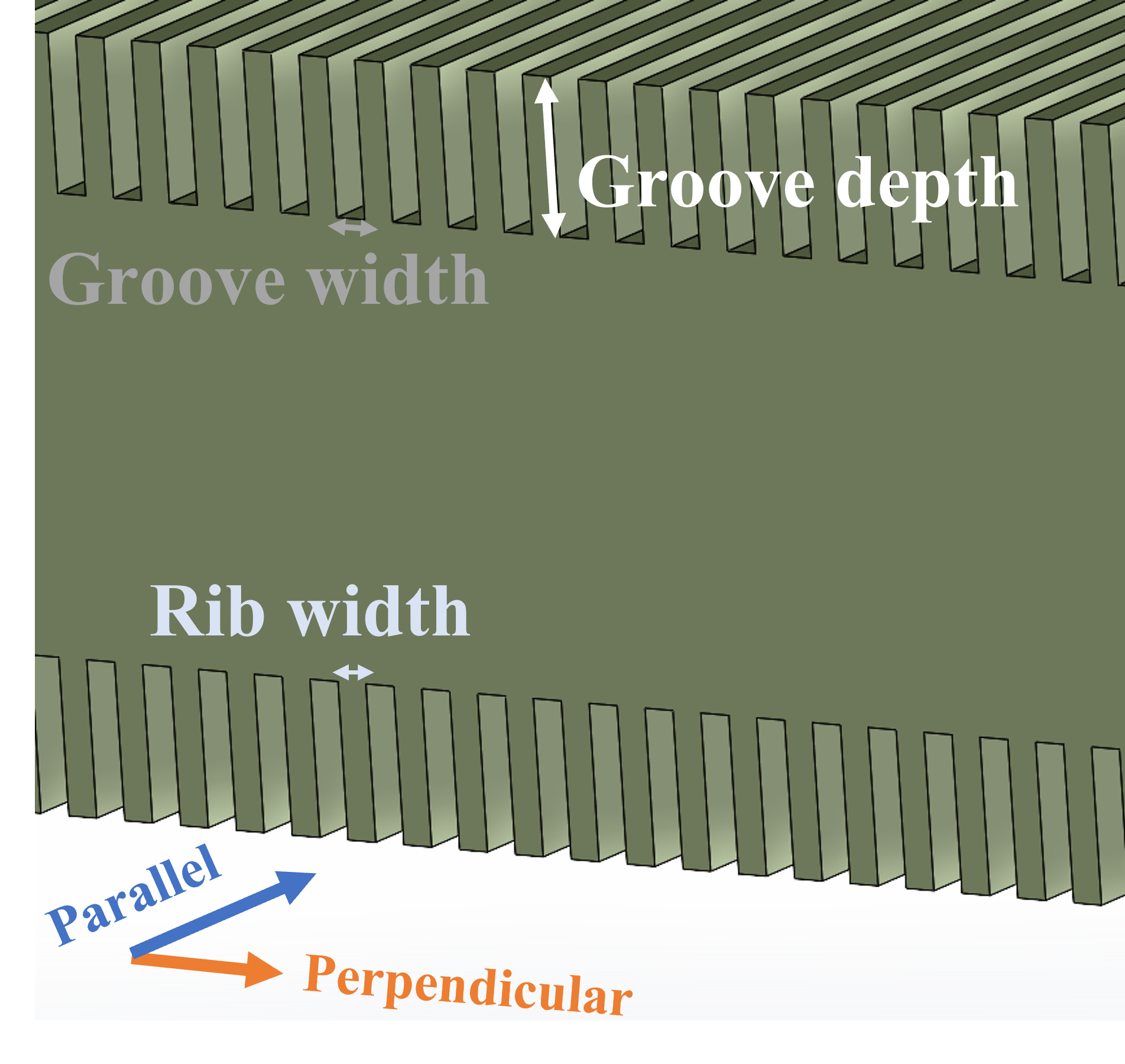}
\qquad
\includegraphics[width=.5\textwidth]{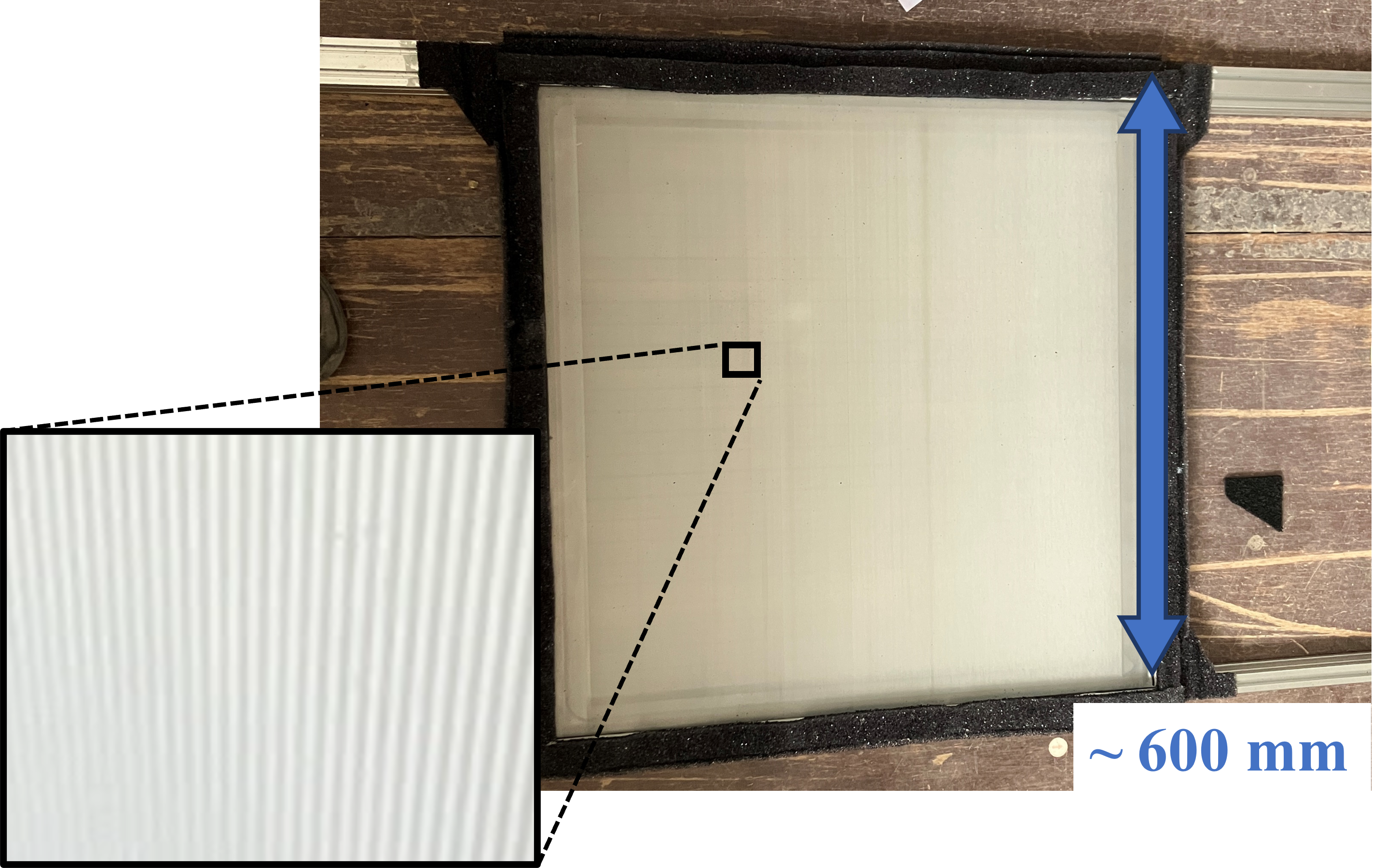}
\caption{QWP drawing and photo. Left: 3D CAD model of the QWP. Right: Photo of the QWP with mounting jigs.\label{fig:QWP_photo}}
\end{figure}

There are two common ways to make a QWP~\cite{vanVliet1981}, and the technique has been well-established. One approach is to fabricate the QWP from a birefringent material, which has different dielectric constants along two orthogonal directions~\cite{MarroneRao_SMA_quartz,Marrone2006_quartz}.
The other approach is to artificially produce the anisotropy by making many grooves in a slab of isotropic material~\cite{Rao1998_BIMA,Rao2006_rexolite}.
We adopt the latter. Our QWP is made of Rexolite~1422, which is cross-linked polystyrene. It is a dielectric with low loss and high transmittance at millimeter-wavelength. The dielectric constant is 2.54~\cite{Lamb}. It is not a birefringent material, but it works as a QWP by cutting grooves (figure~\ref{fig:QWP_photo}).

The Rexolite plate used in SA is \SI{610}{mm} $\times$ \SI{610}{mm} in size. It is determined to fit the existing telescope's structure (see section~\ref{subsec:implimentation} for more details).
\edit{We set the rib and groove widths to be equal. Under this assumption,} the refractive indices parallel (perpendicular) to the grooves, $n_{par}$ ($n_{perp}$), are described as~\cite{Kirschbaum1957_qwp},
\begin{align}
    n_{par}^2 &= \frac{\kappa_1 + \kappa_2}{2} \\
    n_{perp}^2 &= \frac{2\kappa_1 \kappa_2}{\kappa_1 + \kappa_2},
\label{eq:n_qwp}
\end{align}
where $\kappa_1$ and $\kappa_2$ are the dielectric constants of the ribs and the grooves, respectively. \edit{This equal-width configuration approximately maximizes the difference between $n_{par}$ and $n_{perp}$} (figure~\ref{fig:QWP_photo}).
Because the Rexolite dielectric constant is 2.54~\cite{Lamb}, and the dielectric constant of the air is one, $\kappa_2=1$, the design indices are $n_{par}=1.33$ and $n_{perp}=1.20$.
The groove depth is determined to produce a quarter wavelength difference in path length parallel and perpendicular to the grooves. The \edit{initial plate thickness before machining the grooves} is about \SI{15.6}{mm}. Our target frequency is \SI{90}{GHz}, and the calculated design groove depth is \SI{3.18}{mm}.
The groove width and the rib width are \SI{0.635}{mm}. The grooves are cut into the Rexolite plate using a slitting saw on a computer-controlled milling machine. Note that we do not apply an AR coating to the QWP because Rexolite has a relatively low dielectric constant.

To characterize the QWP and express the polarization state, we utilize a Mueller matrix $M$, which is a $4 \times 4$ matrix operating as
$\vec{S}_\mathrm{out} = M \vec{S}_\mathrm{in}$.
Here, $\vec{S}_\mathrm{in}$ ($\vec{S}_\mathrm{out}$) is the input (output) Stokes vectors, $\vec{S}=(I,Q,U,V)^T$.
Our QWP can be modeled as
\begin{equation}
    M_\mathrm{QWP} =
    \begin{pmatrix}
    T & \rho & 0 & 0 \\
    \rho& T & 0 & 0  \\
    0   & 0 & \epsilon & -\eta \\
    0   & 0 & \eta & \epsilon
    \end{pmatrix},
    \label{eq:mueller}
\end{equation}
where $T$ is the transmission, $\rho$ is the differential transmission between the two QWP axes, $\eta$ corresponds to the conversion efficiency, \edit{and $\epsilon$ represents the unconverted Stokes $U$ or $V$ component. In our circular polarization measurement, a nonzero $\epsilon$ appears as U-to-V leakage. An ideal QWP has $T=1$, $\rho=0$, $\eta=1$, and $\epsilon=0$.}

\subsubsection{In-lab characterization}
\label{subsec:design and lab measurements}
We evaluated the QWP by measuring the transmission spectrum in the laboratory.
Both the amplitude and phase of $S_{21}$ were measured over the frequency range from 70 to \SI{114}{GHz} in \SI{30}{MHz} steps using a transmitter and a heterodyne receiver. An optical bench illuminates a spot of about \SI{20}{mm} in diameter on the QWP. The measurements are carried out for electric fields oriented parallel and perpendicular to the grooves.
See figure~9 of Jeong et al.~\cite{Jeong:23} for a more detailed description of the measurement setup.
The measurements made at two widely separated locations on the QWP differ by less than 1\% in amplitude and 3 degrees in phase.

\begin{figure}[tbp]
\centering
\includegraphics[width=.47\textwidth]{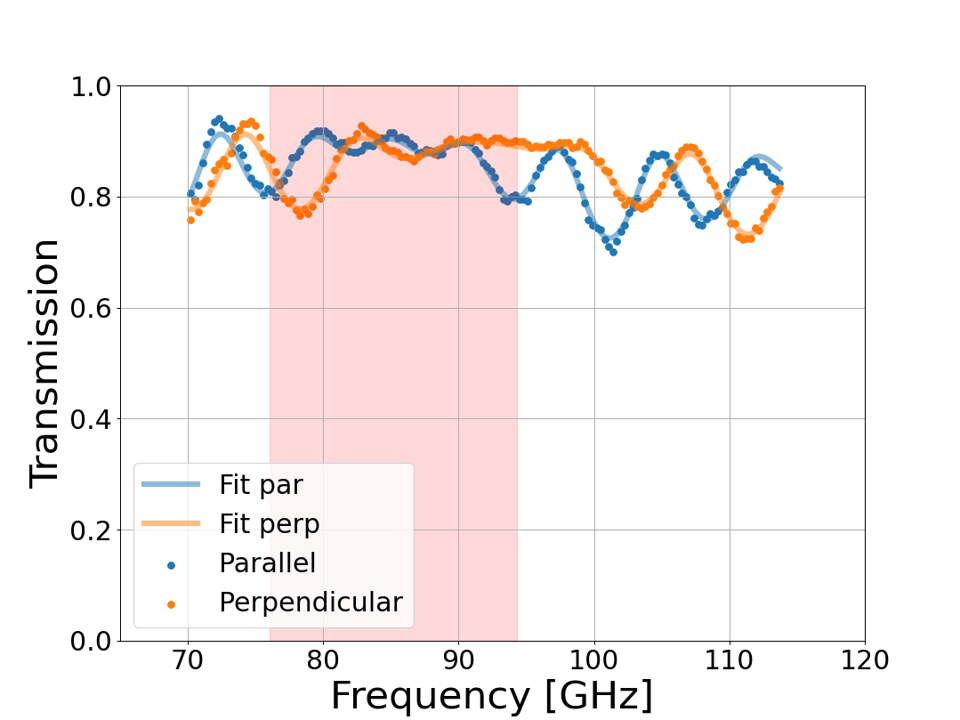}
\qquad
\includegraphics[width=.47\textwidth]{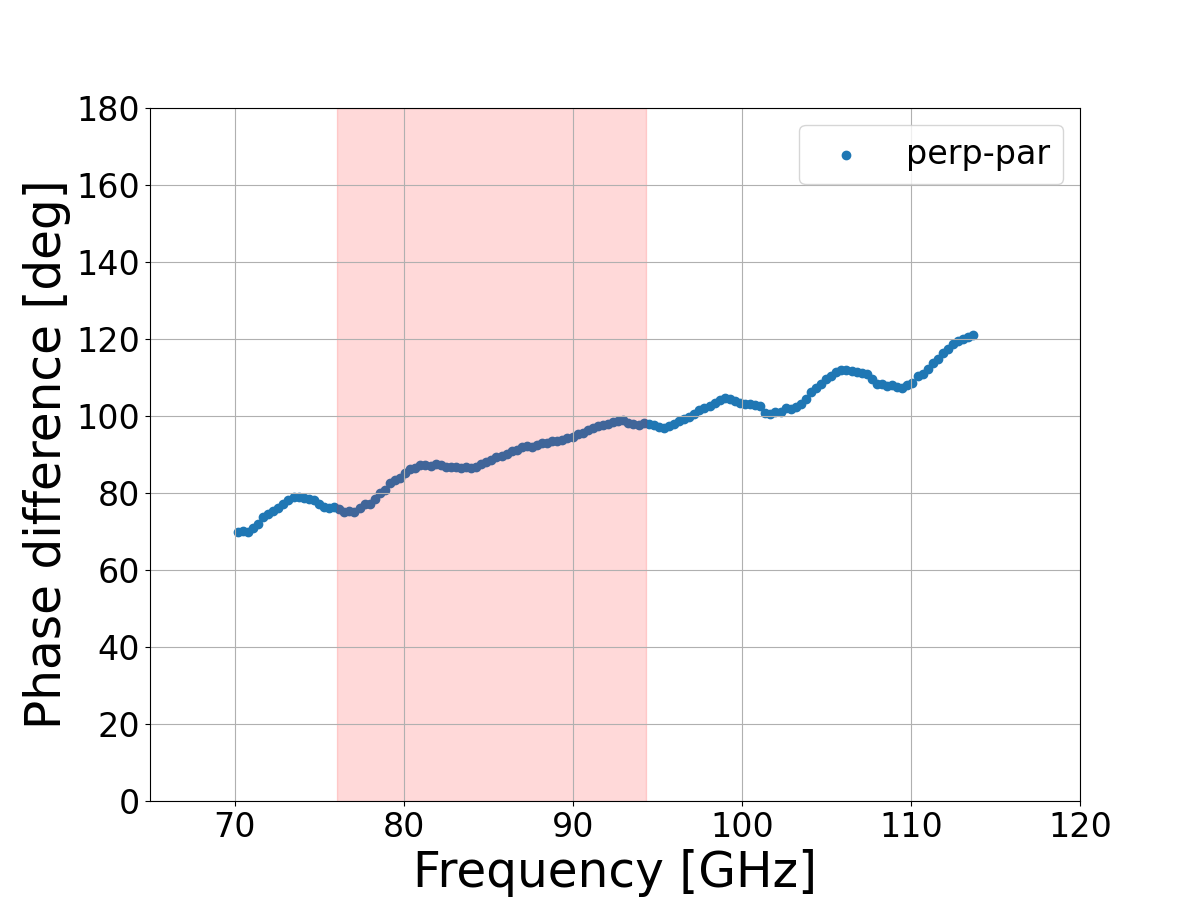}
\caption{QWP measurement results. Left: QWP transmission. The blue dots show the transmission parallel to the grooves, and the orange dots show that perpendicular to the grooves. The lines show the fit results of these transmissions with the thicknesses and the refractive indices using a multilayer dielectric model. The red region shows our \SI{90}{GHz} bandpass. The band center is \SI{85.18}{GHz} and the bandwidth is \SI{18.25}{GHz}. Right: Phase difference in degrees. The QWP provides the required 90-degree phase difference at around \SI{90}{GHz}. \label{fig:QWP_measurement}}
\end{figure}

Figure~\ref{fig:QWP_measurement} shows the lab measurement results of the QWP. The QWP provides the required 90-degree phase difference at around \SI{90}{GHz}. 
The averaged transmission within the bandpass is about 87.8\% and 87.1\% for the \edit{polarizations parallel and perpendicular to the grooves}, respectively. \edit{This corresponds to an estimated thermal emission of about \SI{30}{K} from the QWP, which does not saturate the PB-2b detectors.}
The bandpass is determined based on the PB-2a Fourier-transform spectrometer measurement at a lab. The band center is \SI{85.18}{GHz}, and the bandwidth is \SI{18.25}{GHz} (76.1--94.3\,GHz).

\begin{figure}[tbp]
\centering
\includegraphics[width=.45\textwidth]{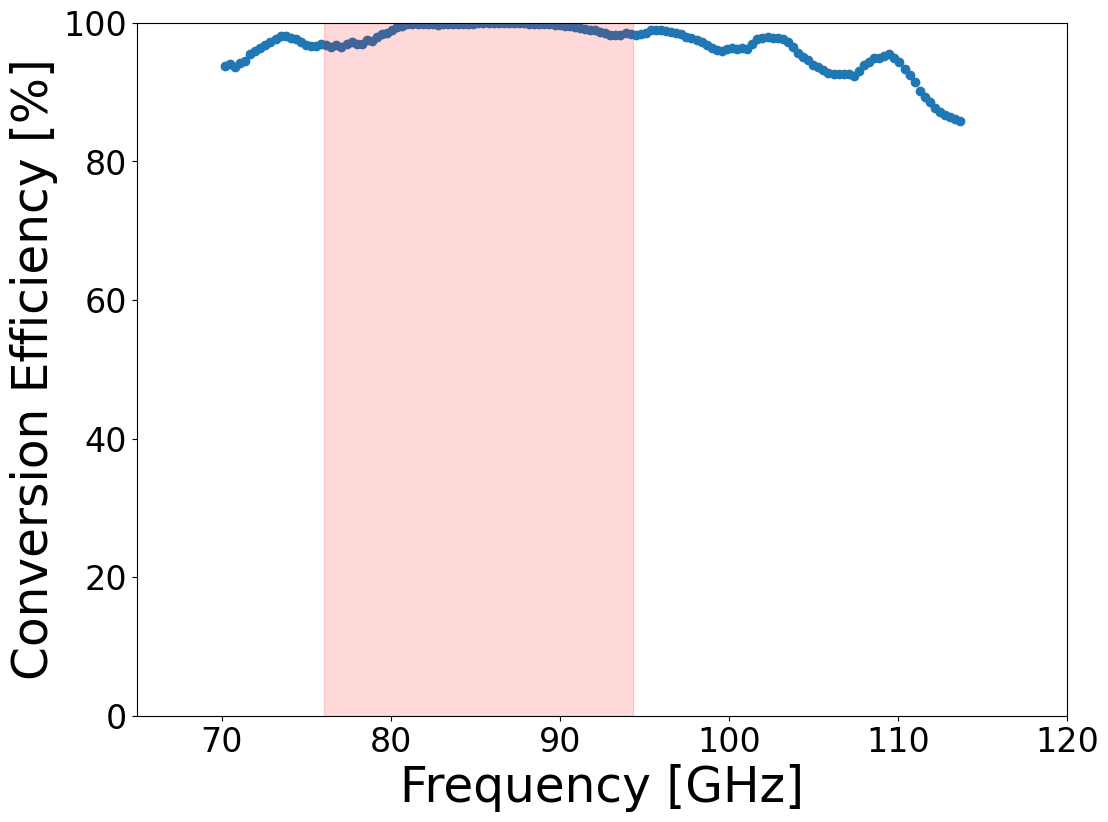}
\qquad
\includegraphics[width=.45\textwidth]{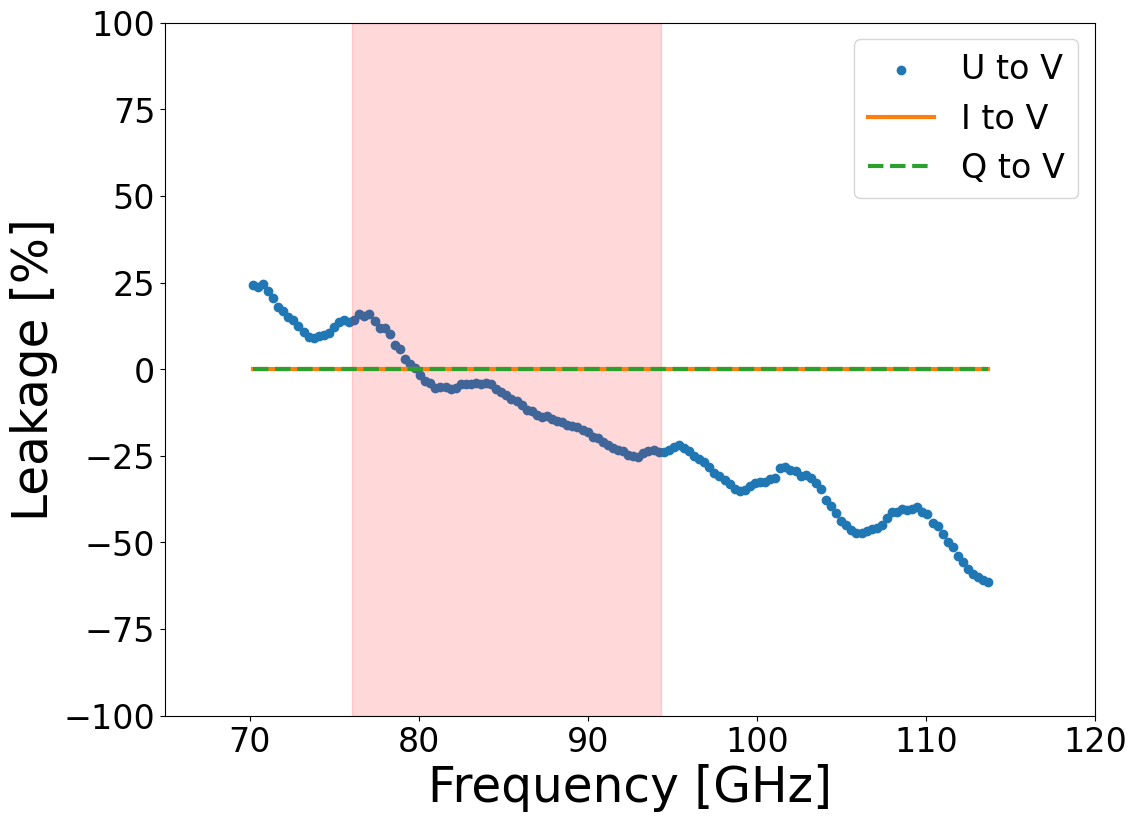}
\caption{Laboratory QWP performance. Left: QWP efficiency. The red region shows our bandpass, the same as figure~\ref{fig:QWP_measurement}. Band-averaged efficiency is 99.0\%. Right: QWP leakage. In this measurement, only U-to-V leakage has non-zero values by definition. 
The band-averaged leakage is $-7.2$\%.
\label{fig:QWP_performance}}
\end{figure}

From the measured transmission and the phase, we estimate the conversion efficiency and the leakage utilizing the Mueller matrix (Figure~\ref{fig:QWP_performance}).
The averaged conversion efficiency, $\eta/T$ in eq.~\eqref{eq:mueller}, is 99.0\%. \edit{The band-averaged laboratory estimate of the U-to-V leakage, $\epsilon/T$, is $-7.2\%$.\footnote{\edit{The U-to-V leakage coefficient used in the analysis is defined as $\epsilon_{UV}=\epsilon/\eta$. Converting the laboratory quantity to this definition requires division by $\eta/T=0.990$, which is negligible at the precision considered here.}} This leakage estimate depends strongly on the adopted bandpass and thus serves only as a reference for comparison with the on-sky characterization.}
\edit{In the QWP Mueller-matrix model of eq.~\eqref{eq:mueller}, there are no I-to-V or Q-to-V leakage terms. These terms can, however, arise from other instrumental effects: I-to-V leakage can result from the combination of QWP differential emission and detector non-linearity, while Q-to-V leakage can result from polarization-angle miscalibration. Together, these leakages into circular polarization can constitute a major source of systematic uncertainty because circular polarization is expected to be small compared with intensity and linear polarization signals. The evaluation of these leakages for the complete observing system, including their on-sky characterization, is discussed in section~\ref{subsubsec:circular polarization}. The effect of U-to-V leakage can also be mitigated through the scan strategy (see Paper~II).}

\subsection{Integration of the QWP into SA}
\label{subsec:implimentation}
\subsubsection{Mounting}
\label{subsubsec:mounting}
We need to design the QWP structure to fit into the existing telescope structure and not interfere with other optics, since the SA is not originally designed for measuring circular polarization.
We install the ambient-temperature QWP on the sky side of the PB-2b receiver, which also means the sky side of the HWP.
It must be large enough to cover all the light rays at the position. On the other hand, it must be small enough not to overlap the optical path between the primary and secondary mirrors (figure~\ref{fig:QWP overview schematics}).
We set the QWP position at a distance of \SI{150}{mm} from the receiver, and we decided on the QWP size requirement.
The aperture size at the QWP position is \SI{430}{mm}. We provide a \SI{50}{mm} margin for diffraction effects and alignment tolerances, and we set a clear aperture requirement $>$~\SI{480}{mm}~\cite{Hill2020,Hill2016SPIE}.
A \SI{610}{mm}$\times$\SI{610}{mm} (\SI{24}{in}$\times$\SI{24}{in}) Rexolite plate, a standard available size, satisfies this requirement and does not obstruct the optical path between the primary and secondary mirrors.

\begin{figure}[tbp]
\centering
\includegraphics[width=.6\textwidth]{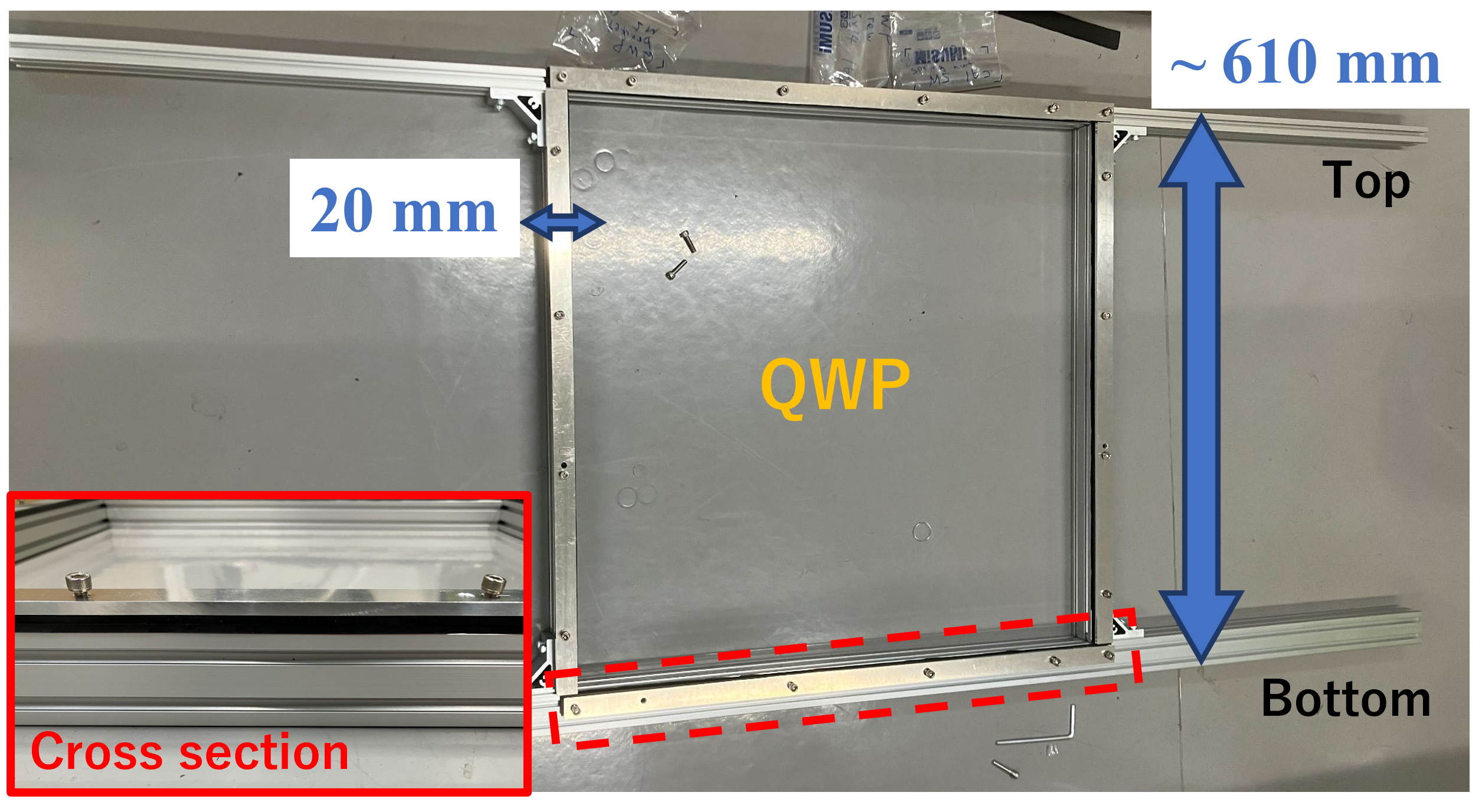}
\qquad
\includegraphics[width=.3\textwidth]{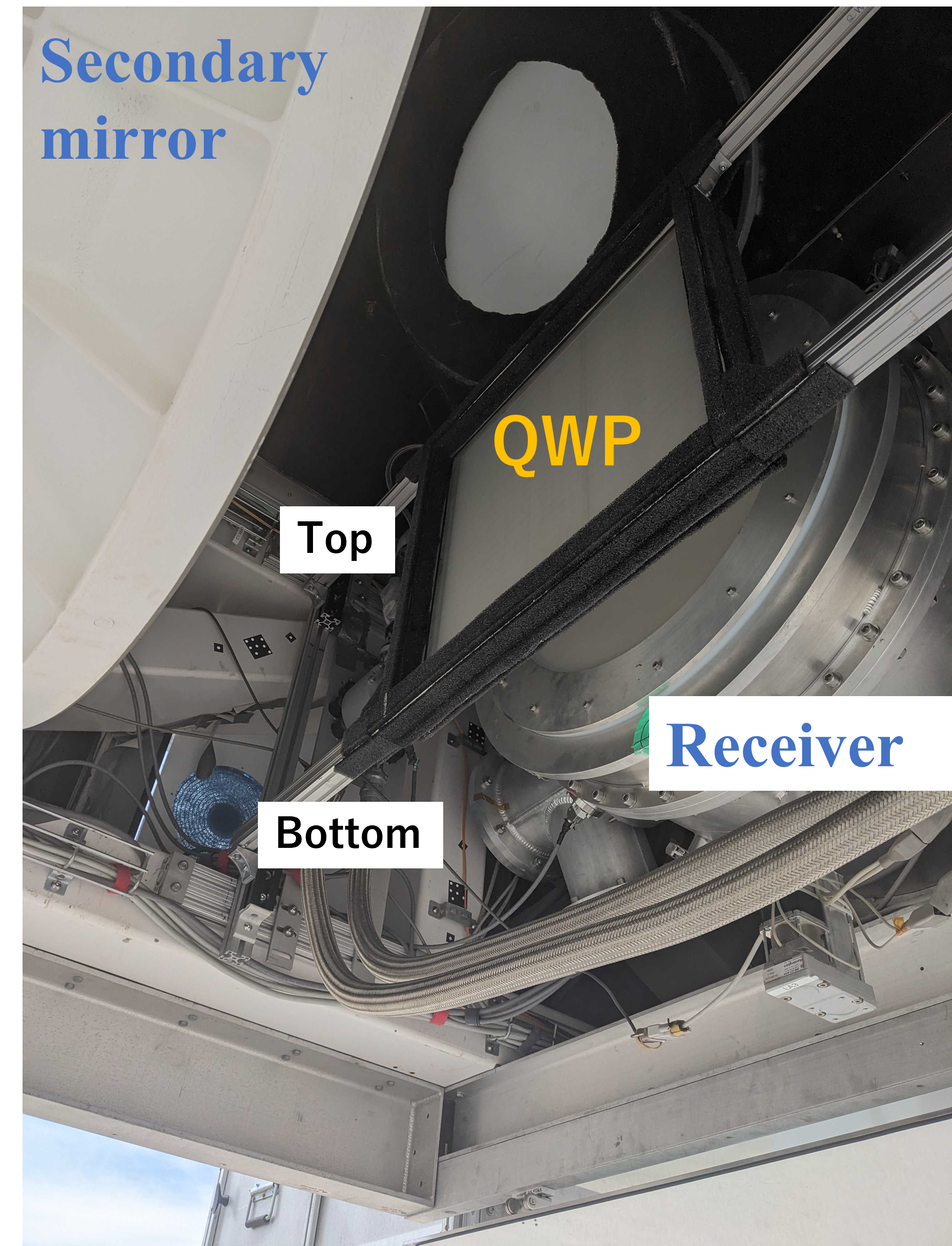}
\caption{Left: Aluminum frames. The photo is without the QWP. The QWP will be mounted to the square region of the photo. Two long aluminum frames extending horizontally are attached to the telescope. The red box shows the cross-section view of the frame. We sandwich the QWP between two \SI{3}{mm}-thick Neoprene sheets (black parts in the view). The width to hold the QWP is \SI{20}{mm} for each side. Right: Photo of the QWP installed.\label{fig:QWP_install}}
\end{figure}

Figure~\ref{fig:QWP_install} (left) shows the aluminum jig to hold the QWP. The long aluminum frames of the jig are attached to the telescope (called ``boom''). The QWP has some holes on all four sides for screws, and it is held in the square region of the jig. The frames to mount the QWP are \SI{20}{mm} around the perimeter, which means the QWP optical surface is \SI{570}{mm} squared. It still matches our aperture requirement. We sandwich the QWP between two \SI{3}{mm}-thick Neoprene sheets\footnote{Neoprene:~\url{https://www.monotaro.com/g/00278067/}} to prevent the plate from breaking.
Figure~\ref{fig:QWP_install} (right) shows the QWP installed at the site. \edit{The QWP grooves are oriented vertically and are aligned with the symmetry plane of the telescope to within \SI{0.3}{degree}.} To prevent the scattering of millimeter waves, aluminum frames are covered with an HR~10 absorber.\footnote{Eccosorb:~\url{https://www.laird.com/}} The QWP does not interfere with any telescope structure. \edit{The entire QWP structure is covered by the receiver enclosure, which provides protection against sunlight and exposure to water.}


\subsubsection{Polarization optics model}
\label{subsubsec:pol_opt_model}
\begin{figure}[tbp]
    \centering
    \includegraphics[width=1\linewidth, keepaspectratio]{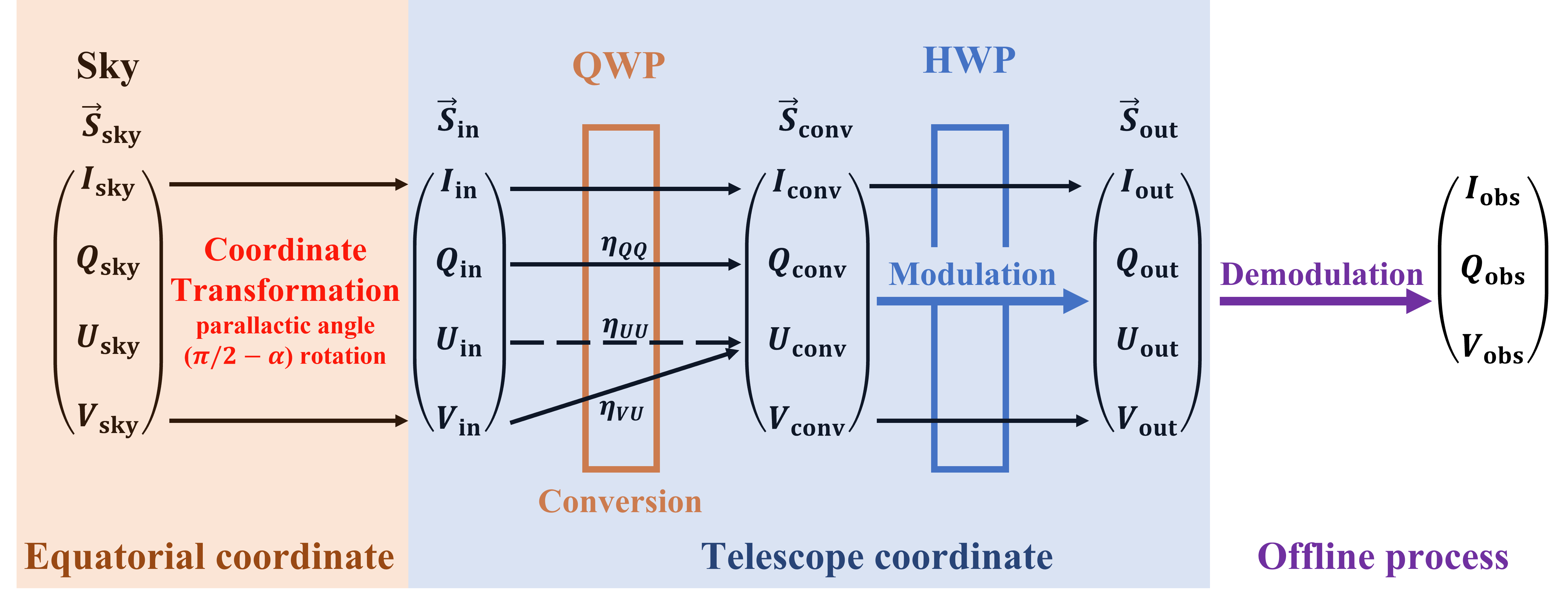}
    \caption{Schematic diagram of the polarization signal flow and transformations. Sky signals propagate from left to right. \edit{$\vec{S}_\textrm{sky}$, $\vec{S}_\textrm{in}$, $\vec{S}_\textrm{conv}$, and $\vec{S}_\textrm{out}$ denote the Stokes vector $(I,Q,U,V)^T$ in sky coordinates, in telescope coordinates before the QWP, after the QWP, and after the HWP, respectively. The quantities $I_\textrm{obs}$, $Q_\textrm{obs}$, and $V_\textrm{obs}$ are reconstructed after demodulation.} The telescope coordinate \edit{system} is fixed to the QWP. This coordinate system is related to equatorial coordinates through the parallactic angle $\alpha$ of observations. $\eta_{ij}$ is the efficiency with which incoming $i$ is converted to $j$ by the QWP. In the ideal case, $\eta_{QQ}=\eta_{VU}=1$ and $\eta_{UU}=0$. \edit{The HWP spins at \SI{2}{Hz} and modulates linear polarization.} The process on the receiver side of the QWP (i.e., the right-hand side in the figure) is consistent with that of PB-2a. Incoming circular polarization is obtained by demodulation in an offline process.}
    \label{fig:optics schem demod}
\end{figure}

\edit{Figure~\ref{fig:optics schem demod} shows the polarization signal flow through the optical system and defines the Stokes-parameter notation used in this paper. The Stokes vectors $\vec{S}_\textrm{sky}$, $\vec{S}_\textrm{in}$, $\vec{S}_\textrm{conv}$, and $\vec{S}_\textrm{out}$ denote $(I,Q,U,V)^T$ in sky coordinates, in telescope coordinates before the QWP, after the QWP, and after the HWP, respectively. After demodulation, the reconstructed quantities are denoted by $I_\textrm{obs}$, $Q_\textrm{obs}$, and $V_\textrm{obs}$.}

\edit{The reconstructed circular polarization signal, $V_\textrm{obs}$, includes the input circular polarization attenuated by the efficiency, together with leakage due to imperfections of the telescope system. We write the relevant response as
\begin{equation}
    V_\textrm{obs}
    = M_{VV} \, V_\textrm{in}
    + M_{IV} \, I_\textrm{in}
    + M_{UV} \, U_\textrm{in}
    + M_{QV} \, Q_\textrm{in},
    \label{eq:V_response}
\end{equation}
where $M_{VV}$ represents the circular polarization efficiency and the other three coefficients characterize leakage into $V_\textrm{obs}$.
Similarly, we define $M_{II}$ and $M_{QQ}$ as the responses from $I_\textrm{in}$ to $I_\textrm{obs}$ and from $Q_\textrm{in}$ to $Q_\textrm{obs}$, respectively. We then define the corresponding leakage coefficients used in the analysis as\footnote{The different normalizations are intentional and follow those used in the analysis: the I-to-V leakage is characterized relative to the intensity response, whereas the U-to-V and Q-to-V leakages are characterized relative to the circular-polarization response.}
\begin{equation}
    \epsilon_{IV}\equiv\frac{M_{IV}}{M_{II}}\:,\qquad
    \epsilon_{UV}\equiv\frac{M_{UV}}{M_{VV}}\:,\qquad
    \epsilon_{QV}\equiv\frac{M_{QV}}{M_{VV}}\:.
\end{equation}
}

\edit{The circular polarization efficiency $M_{VV}$ can be related to the linear polarization efficiency, $M_{QQ}$, which is particularly useful because a celestial calibrator with strong, well-known circular polarization is not available. Here, we assume that the efficiencies for Stokes $Q$ and $V$ differ only by the QWP response. For the telescope optics, the cross-polarization is sufficiently small, and thus the telescope response can be approximated as a common efficiency factor for all the Stokes parameters. For the receiver, the cryostat optics are approximately rotationally symmetric and the detector polarization-angle distribution has no strong directional bias. Thus, after averaging over detectors, the polarization efficiency can be approximated as being the same for Stokes $Q$ and $U$. Denoting the polarization efficiency independent of the QWP by $\eta_P$, the efficiencies for $V$ and $Q$ can then be written as
\begin{equation}
\begin{aligned}
    M_{VV}\equiv\eta_V&=\eta_{VU}\,\eta_P \:, \\
    M_{QQ}\equiv\eta_Q&=\eta_{QQ}\,\eta_P \:, 
\end{aligned}
\label{eq:quv_efficiency}
\end{equation}
where $\eta_{ij}$ is the efficiency with which Stokes $i$ is converted into Stokes $j$ by the QWP. In terms of the QWP Mueller-matrix elements defined in eq.~\eqref{eq:mueller}, $\eta_{QQ}=T$ and $\eta_{VU}=\eta$. Thus, $M_{VV}$ can be determined by combining the on-sky calibration of $M_{QQ}$ with the laboratory measurement of the QWP response:
\begin{equation}
    M_{VV}=M_{QQ}\,\frac{\eta_{VU}}{\eta_{QQ}}
          =M_{QQ}\,\frac{\eta}{T} \:.
          \label{eq:mvv_mqq_relation}
\end{equation}}

\edit{The U-to-V response, $M_{UV}$, is characterized on sky using Tau~A. The non-ideal QWP response is expected to dominate this term. Neglecting other contributions, it can be written as
\begin{equation}
    M_{UV}\equiv \eta_U =\eta_{UU}\,\eta_P,
\end{equation}
where $\eta_{UU}=\epsilon$ in eq.~\eqref{eq:mueller}. The corresponding U-to-V leakage is therefore
\begin{equation}
    \epsilon_{UV}=\frac{M_{UV}}{M_{VV}}
    =\frac{\eta_{UU}}{\eta_{VU}}
    =\frac{\epsilon}{\eta} \:.
    \label{eq:UV_leakage}
\end{equation}
This relation allows the laboratory QWP measurement to provide an estimate of the U-to-V leakage for comparison with the on-sky characterization.}

\edit{The I-to-V response, $M_{IV}$, is not present in the QWP Mueller matrix of eq.~\eqref{eq:mueller}, but can result from other instrumental effects, including detector non-linearity. This term is corrected on an observation-by-observation basis, and its main impact on the final measurement is the uncertainty associated with the correction (sections~\ref{subsec:data processing} and~\ref{subsec:I-to-V uncertainty}). Similarly, the QWP Mueller matrix does not contain a Q-to-V term. Given the small telescope cross-polarization, the relevant $M_{QV}$ contribution is dominated by polarization-angle miscalibration in the receiver. This effect is subdominant compared to the other systematics (section~\ref{subsubsec:circular polarization}).}

\edit{With the $M_{IV}$ term corrected on an observation-by-observation basis and the subdominant $M_{QV}$ term neglected, the observed signal $V_\textrm{obs}$ can be related to the sky signal as
\begin{equation}
    V_\textrm{obs} = \eta_V \, V_\textrm{sky} + \epsilon_{UV} \, \eta_V \, P_\textrm{sky} \, \sin(2\alpha),
\label{eq:Vo_expression}
\end{equation}
where $P_\textrm{sky} \equiv \sqrt{Q_\textrm{sky}^2+U_\textrm{sky}^2}$ is the linear polarization in sky coordinates and $\alpha=\gamma-\alpha_{\rm obs}+\pi/2$ is the angle of the source linear polarization in telescope coordinates. Here, $\gamma$ and $\alpha_{\rm obs}$ are the source polarization angle and the observation's parallactic angle, respectively.
The U-to-V leakage contribution depends on the source parallactic angle, while a sky $V$ signal is constant with parallactic angle. This dependence allows us to characterize the leakage (section~\ref{subsubsec:circular polarization}) as well as to mitigate the leakage through scan strategy (Paper~II).}

\edit{For the remainder of this paper, we adopt the overall response normalization such that the Stokes $I$ efficiency of the system excluding the QWP is unity. With this convention, $M_{II}=\eta_{II}=T$. This choice does not affect the quantities considered in this analysis or those reported in Paper~II, since they are constructed from ratios of Stokes signals or response coefficients, such as polarization fractions and leakage coefficients, for which a common overall normalization cancels. Accordingly, $\eta_P$ represents the polarization efficiency of the system excluding the QWP relative to its Stokes $I$ efficiency. The source circular-polarization fraction reported in this paper and Paper~II is expressed in terms of the equivalent sky signal; after accounting for the leakage terms in the measured ratio $V_\textrm{obs}/I_\textrm{obs}$, the resulting ratio is divided by the relative response $M_{VV}/M_{II}=\eta\,\eta_P/T$.}

\section{Dataset}
\label{subsec:dataset}

We installed the QWP in December 2023, and PB-2b observations were conducted from December 2023 to September 2024. 
Tau~A and Jupiter \edit{were} observed for calibration during \edit{this} period.
Because the observed circular polarization can be written as the sum of the source circular polarization and leakages (equation~\eqref{eq:Vo_expression}), observations with multiple parallactic angles allow us to calibrate the \edit{U-to-V} leakage and \edit{polarization} efficiency.

Observation conditions are similar to PB-2a scans. First, we apply the Sun and Moon avoidance. \edit{We observed point sources when their positions were more than 15 degrees from the Sun and Moon to avoid contamination through the sidelobe.} Then, we operate the telescope at an elevation between 30 and 60 degrees. It is high enough not to pick up signals from the ground and not too high compared to the elevation during tuning, which is approximately 40 degrees.

Some of the radio point sources are observed both when they are rising and setting, which means we observe them with multiple parallactic angles.
One observation is about 1 to 1.5 hours long.
The scan pattern is called a raster scan. It scans in azimuth back and forth while tracking in elevation. One azimuth sweep is referred to as a subscan. The duration of one subscan is about 15 seconds. The scan speed is \SI{0.4}{deg/sec} on the sky.

\edit{A total of 114 Tau~A observations were conducted.} From that data, we cut observations with known instrumental issues, bad precipitable water vapor (PWV) conditions, or the number of detectors $< 20$ after data selection (see section~\ref{sec:analysis}). The final number of observations is 19, which is about 17\% of the total observations.
Jupiter was continuously observed for commissioning purposes since first light. The number of Jupiter observations with the QWP installed is 48. Applying the same observation selection criteria as Tau~A, we use 10 observations (21\%) for this analysis.


In addition to Tau A and Jupiter, we observe seven radio point sources, which are RCW~38, IRAS 16183-4958, IRAS 19078+0901, RCW~122, Cen~A, 3C~279, and 3C~273. These results are presented in Paper II. 

\section{Analysis}
\label{sec:analysis}
Tau~A and Jupiter datasets are analyzed to validate the overall pipeline and evaluate systematic effects.
Since we have the HWP rotating at \SI{2}{Hz}, the raw time-ordered data (TOD) is modulated at \SI{8}{Hz}, four times the HWP rotation frequency. We reconstruct the intensity and polarization by demodulation. 
After calibrating the TOD, we make flat sky maps to estimate the circular polarization fraction. 
We calculate the fraction of the observed Stokes $V_\textrm{obs}$ to the intensity $I_\textrm{obs}$ of the sources.
The overall pipeline is similar to those used for linear polarization analysis, PB-2a Tau~A analysis, for example~\cite{2aTauA}.

\subsection{Data processing}
\label{subsec:data processing}
The rotating HWP modulates polarization signals in the detector TOD. The measured modulated signal of a single detector is written as
\begin{equation}
    d_m = I_\textrm{conv} + \eta_P \textrm{Re}[{(Q_\textrm{conv}+iU_\textrm{conv})m(\chi)e^{i2\theta_\textrm{det}}}]+A(\chi)+N_w,
    \label{eq:modulated TOD}
\end{equation}
where $I_\textrm{conv}$, $Q_\textrm{conv}$, and $U_\textrm{conv}$ are the incoming Stokes parameters to the HWP~\cite{Kusakahwp,Takakura_2017}. As shown in figure~\ref{fig:optics schem demod}, they correspond to the Stokes parameters after conversion by the QWP. Here, $\chi$ is the HWP angle, $\eta_P$ is the linear polarization efficiency of the entire system excluding that of the QWP (section~\ref{subsubsec:pol_opt_model}), $m(\chi)=\exp(-i4\chi)$ is the modulation function, and $\theta_\textrm{det}$ is the detector angle in telescope coordinates.
$A(\chi)$ is the spurious modulation signal that correlates with the HWP angle, which is referred to as the HWP synchronous signals (HWPSSs). $N_w$ is the detector white noise~\cite{Kusakahwp}.
\edit{The HWPSS is removed from the TOD as a contaminant, while its phase is also used to calibrate the detector orientation relative to the QWP (section~\ref{subsec:characterization and calibration}).}

The HWPSS can be expressed by two components, one of which is independent of the sky intensity and the other which is proportional to it, as
\begin{equation}
    A (\chi) = A_0 (\chi) + \lambda (\chi) \, \edit{I_\textrm{obs}}.
\label{eq:hwpss}
\end{equation}
Each component is decomposed as
\begin{align}
    A_0(\chi) &= \sum_n(A_n'\cos(n\chi)+B_n'\sin(n\chi)) \\
    \lambda(\chi) &= \sum_n(\lambda_n^r\cos(n\chi)+\lambda_n^i\sin(n\chi)), \label{eq:hwpss leakage}
\end{align}
where subscript $n$ denotes the $n$th harmonic and $\chi$ again is the HWP rotation angle.
The optical elements of the sky side of the HWP generate the HWPSS~\cite{Takakura_2017,Kusakahwp}. Since we have off-axis mirrors and the birefringent QWP, the fourth harmonic of HWPSS ($4f$ HWPSS), which is aligned to the QWP and mirror directions, is dominant.
We subtract the HWPSS in TOD by calculating the HWPSS amplitude vs. the HWP angle. First, we make an HWPSS model from TOD in each detector. Then, we reconstruct the time vs. HWPSS model using the HWP encoder angle. Finally, we subtract the HWPSS model from the timestreams. This HWPSS is calculated in each subscan considering the HWP fluctuation. The typical HWPSS temperature is found to be about \SI{2}{K}. The TOD scanning over the source region is masked when making the HWPSS model, with a circular mask of diameter 0.5 deg around the source.

We demodulate the modulated signal, $d_m$, to reconstruct the incoming signals to the HWP. The intensity is reconstructed by applying lowpass filtering.
We apply the filter at \SI{3.8}{Hz} to avoid picking up a significant $4f$ HWPSS described later in this section.
To obtain the polarization signals, we apply a bandpass filter around the modulation frequency of \SI{8}{Hz} and then demodulate the TOD by multiplying the complex conjugate of the modulation function $m(\chi)$.
We then apply a lowpass filter to the demodulated TOD. The real and imaginary parts of the lowpass-filtered TOD correspond to the incoming Stokes $Q$ and $U$ parameters, $Q_{\rm conv}$ and $U_{\rm conv}$, respectively. \edit{The reconstructed polarization signals, $Q_\textrm{obs}$ and $V_\textrm{obs}$, can be written in terms of $Q_\textrm{in}$, $U_\textrm{in}$, and $V_\textrm{in}$ as described in section~\ref{subsubsec:pol_opt_model}:}
\begin{equation}
\begin{aligned}
    Q_\textrm{obs}&\simeq\eta_{Q}\,Q_\textrm{in} \\
    V_\textrm{obs}&\simeq\eta_{V}\,V_\textrm{in}+\eta_{U}\,U_\textrm{in}.
\end{aligned}
\end{equation}

The leakage in the demodulated signal proportional to the unpolarized intensity is so-called intensity-to-polarization (I-to-P) leakage.
There are two kinds of leakage relevant to this analysis, ``I-to-Q leakage'' and ``I-to-V leakage.'' The I-to-Q leakage is the leakage to the $Q$ signal in the telescope frame. It is caused by imperfections of optical elements like the QWP and non-linearity of the detectors~\cite{Takakura_2017}. On the other hand, the I-to-V leakage can arise from the non-linearity of the time constant of detectors. The time constant has temporal variations during the observation. The variations proportional to the incoming signal fluctuations generate an imaginary component in the demodulated signals~\cite{Takakura_2017}. The QWP itself does not make the I-to-V leakage~(section~\ref{subsec:mechanism}).
We subtract the leakage in timestreams. We follow the PB-2a Tau~A analysis method~\cite{2aTauA}, \edit{which uses low-frequency atmospheric fluctuations to estimate the leakage coefficients~\cite{Takakura_2017,Johnson_2007}.
We calculate the I-to-Q and I-to-V leakage coefficients, $\lambda_4^r$ and $\lambda_4^i$, respectively, from the correlation between the low-frequency fluctuations in the intensity and polarization timestreams~\cite{Takakura_2017}.  Here, $\lambda_4^i$ corresponds to the I-to-V leakage coefficient $\epsilon_{IV}$ defined in section~\ref{subsubsec:pol_opt_model}.}
The coefficients are calculated in the time domain after lowpass filtering at \SI{0.4}{Hz} per subscan. Further discussion of systematic uncertainties due to
this leakage is in section~\ref{subsec:I-to-V uncertainty}.


We apply polynomial filters to the demodulated TOD, to remove the long-timescale fluctuations of the atmosphere and instruments. We apply a third-order polynomial filter to both intensity and polarization in each subscan, with the source masked out \edit{using the same 0.5-degree-diameter circular mask}~\cite{Adachi_2020,Ade_2014}.

\subsection{Data selection}
\label{subsec:Data selection}
We apply two types of data selections, ``observation selection'' and ``detector selection.'' 
Regarding observation selections, we choose the observations based on the specific observation conditions. The criteria are similar to those of PB-1 Tau~A and PB-2a Tau~A analysis~\cite{Yamada2024,2aTauA}.
After excluding observations with known instrumental problems, we select observations with favorable weather conditions. The criteria are that the average PWV $< 4$\,mm and (Max PWV - Min PWV) $<$ \SI{1}{mm} over the course of the observation. The PWV data comes from the Atacama Pathfinder EXperiment (APEX) radiometer~\cite{APEX2006}. It is located about \SI{5}{km} southeast of the SA telescopes.
Additionally, we select observations that have more than 20 detectors after the detector selections described below.
The final number of Tau~A observations that meet these selection criteria is 19.

The detector selection is divided into two parts. The first part is the selection common to observations. Of all PB-2b's detectors, we select those that are sensitive to \SI{90}{GHz} and have pointing offset information~\cite{Russell2024}. The number of these detectors is 1018.

The second part is performed observation by observation. This is run during the data processing, which means that we apply the data selection for each TOD. It is divided into four steps.
The first is the tuning status. We selected the successfully tuned bolometers for each observation.
Next, we verify whether the selected detectors can detect optical signals. We utilize the HWPSS for this selection. To characterize the HWPSS (Equation~\eqref{eq:hwpss}), we fit the HWPSS angle vs. amplitude in TOD as
\begin{equation}
    A_0 = \sum_{n=1}^6 A'_n \sin(n\chi)+B'_n \cos(n\chi),
\end{equation}
where the amplitude of the $n$-th harmonic is $A_n=\sqrt{A_n'^2+B_n'^2}$. The data selection is conducted with the fourth harmonic amplitude. We normalize the $A_4$ by the RMS of fit residuals.
We select detectors that have $>10$ HWPSS SNR.

\begin{figure}[tbp]
    \centering
    \includegraphics[width=1.\linewidth,keepaspectratio]{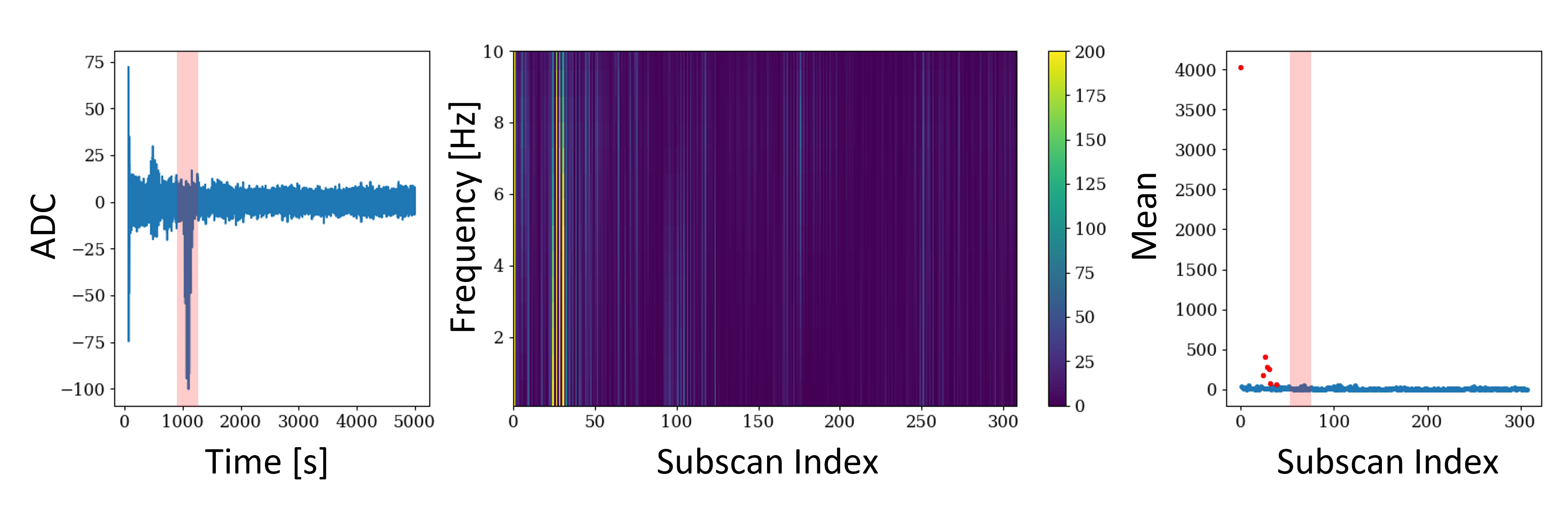}
    \caption{Spectrogram analysis of one detector during a Tau~A observation. Left: Time vs. detector ADC (arbitrary units) after polyfiltering. The peak around \SI{1000}{s} out in the negative direction is a Tau~A signal. The red region indicates the time range related to the selection criteria for this detector. The spikes before the signals are baseline drifts. Large counts at the beginning are due to the telescope movement to the source. Middle: Spectrogram result. We divide the TOD into subscans that last about 15 seconds, and these subscans correspond to the entire observation time in the left figure. The color map shows the power at each frequency in the Fourier domain. We mask the source when calculating the spectrogram. The yellow feature around subscan 25 indicates increased power caused by baseline drift. Right: Mean power in each subscan, calculated using frequencies below 3.8 Hz. Red dots show outliers. There is no outlier in the red region, and this detector passed the selection.}
    \label{fig:spectrogram}
\end{figure}

Then, we select the bolometers based on a spectrogram. A baseline drift occurs randomly due to issues such as detector non-linearity, so we check for no drift around the target source. We calculate the mean power lower than \SI{3.8}{Hz} in the Fourier domain. Figure~\ref{fig:spectrogram} shows an example of the spectrogram analysis. The power is calculated in every subscan, and we define the outlier point as
\begin{equation}
    P_i - \textrm{median}(P_i) > 10 \times \textrm{median}(|P_i - \textrm{median}(P_i)|),
\end{equation}
where $P_i$ is the $i$-th subscan mean power lower than \SI{3.8}{Hz}. The TOD is masked while scanning the source region. We select detectors that show no outliers when the source is hitting. 
This criterion requires the absence of outliers in the subscan containing the source and in the surrounding $\pm5$ subscans.

Finally, we cut detectors based on I-to-P leakage coefficients. The coefficient of I-to-P leakage is calculated as a median of each subscan coefficient. However, some of the coefficients in a single observation exhibit significant variations. To select detectors with stable leakage coefficients, we set the criteria as
\begin{equation}
    \frac{Q3-Q1}{2} < 0.03.
\end{equation}
Here, $Q1$ ($Q3$) represents the first (third) quartile of the leakage coefficients of each detector during an observation, where the leakage is calculated in each subscan.
While this leakage cut is the largest hit to the data selection, it is necessary to limit the uncertainty introduced by the I-to-V leakage correction. This correction is the largest source of systematic uncertainty in the final circular polarization measurement (section~\ref{subsubsec:circular polarization}; see also Paper~II), and the cut is needed to maintain a balance between systematic and statistical uncertainties. \edit{In particular, this selection limits fluctuations in the leakage correction by requiring the leakage coefficient to be stable within each observation, rather than selecting on the measured source polarization.}

Table~\ref{tab:data selection} shows the \edit{cumulative number of detectors after each selection, summed over} all observations that are used for scientific analysis. The final \edit{data-selection efficiency for Tau~A (Jupiter) is 20\% (17\%)}.

\begin{table}[tbp]
\centering
\caption{Cumulative number of detectors by data selection. Values in the parentheses are the percentage of data selection. \edit{The percentages are calculated relative to the cumulative number of tuned \SI{90}{GHz} detectors for each source.} It is the number of operable detectors for which we have pointing data across all 19 (10) observations. The final efficiency of Tau~A (Jupiter) observation is 20\% (17\%).\label{tab:data selection}}
\begin{tabular}{c|c|cccc}
Source          &  No. of obs & \begin{tabular}[c]{@{}c@{}}Tuned \SI{90}{GHz}\\Total\end{tabular} & \begin{tabular}[c]{@{}c@{}}HWPSS SNR\\ ({[}\%{]})\end{tabular} & \begin{tabular}[c]{@{}c@{}}Spectrogram\\ ({[}\%{]})\end{tabular} & \begin{tabular}[c]{@{}c@{}}Leakage coef.\\ ({[}\%{]})\end{tabular}    \\ \hline
Tau A           &  19 & 13068                                                             & 11609 (89)                                                     & 11346 (87)                                                       & 2577 (20) \\
Jupiter & 10 & 7842 & 7495 (96) & 4072 (51) & 1336 (17)
\end{tabular}
\end{table}

\subsection{Calibration}
\label{subsec:characterization and calibration}
There are four types of detectors, each group sensitive to microwaves with different polarization directions, in each wafer on the focal plane. After demodulation, we rotate the detector angle to reconstruct polarizations and obtain $Q_\textrm{obs}$ and $V_\textrm{obs}$ maps on the telescope coordinates.
We find that the HWPSS is significant compared to the sky signal, about \SI{2}{K}, after installing the QWP, and we utilize it for calibration. The HWPSS is aligned to the QWP vertical axis, which is the same as the mirror polarization direction.\footnote{We confirm it by comparing angles between observations with and without the QWP. We also check angles using Tau~A to confirm no obvious angle miscalibration.} The angle of the detectors is calculated using the phase of the fourth harmonic of the HWPSS, which reflects the QWP's physical orientation. It is calculated in each observation considering the time constant variation. Note that this analysis does not need the absolute detector polarization angle.

The detector pointing offsets are calculated from planet scans~\cite{Russell2024}. 
We use this array pointing for all observations to make source-centered maps.

Gain is the detector's responsivity to the input signals. Since we report the circular polarization ratio to intensity, we only calibrate the gain between detectors, which is called relative gain. We use the fourth-harmonic amplitude of the HWPSS, assuming that detectors see the same HWPSS.

One of the biggest systematics of this analysis is the leakage. We developed the leakage subtraction and uncertainty estimation method from Tau~A and Jupiter observations. Details are discussed in section~\ref{subsec:characterization}. 


\subsection{Map making and noise characterization}
\label{subsec:map making}
From the demodulated TOD, we create maps of intensity and polarizations. The mapmaking pipeline is similar to PB-2a analysis~\cite{2aTauA}. Our map is a source-centered map using the detector pointing offset in telescope coordinates.
The map is pixelized to \SI{0.02}{deg.} $\times$ \SI{0.02}{deg.} per pixel.
\edit{The reconstructed $I_\textrm{obs}$ and $V_\textrm{obs}$ signals are projected onto maps and coadded with inverse-noise weights. We denote the resulting coadded intensity and circular-polarization maps by $I_\textrm{coadd}$ and $V_\textrm{coadd}$, respectively.}

\edit{The measured circular-polarization fraction is calculated from these coadded maps by integrating pixels within a 0.1-degree radius as}
\begin{equation}
V_{\textrm{frac}} = \frac{\Sigma_p^{r < 0.1}V_{\textrm{coadd},p}}{\Sigma_p^{r < 0.1}I_{\textrm{coadd},p}},
\label{eq:V_fraction}
\end{equation}
where $I_{\textrm{coadd},p}$ and $V_{\textrm{coadd},p}$ are the $p$th $I_{\textrm{coadd}}$ and $V_{\textrm{coadd}}$ map pixels in the coadded map.
\edit{Thus, $V_\textrm{frac}$ is the map-domain counterpart of $V_\textrm{obs}/I_\textrm{obs}$, obtained after reconstructing $V_\textrm{obs}$ and $I_\textrm{obs}$ into maps and integrating over the source region.}

\begin{figure}[tbp]
    \centering
    \includegraphics[width=0.4\linewidth,keepaspectratio]{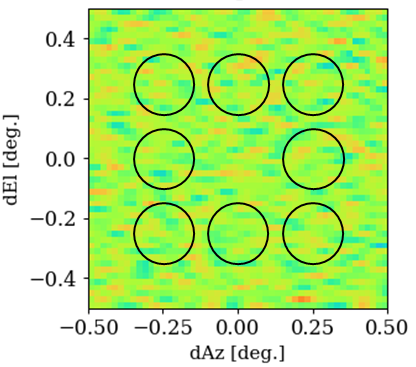}
    \qquad
    \includegraphics[width=0.53\linewidth,keepaspectratio]{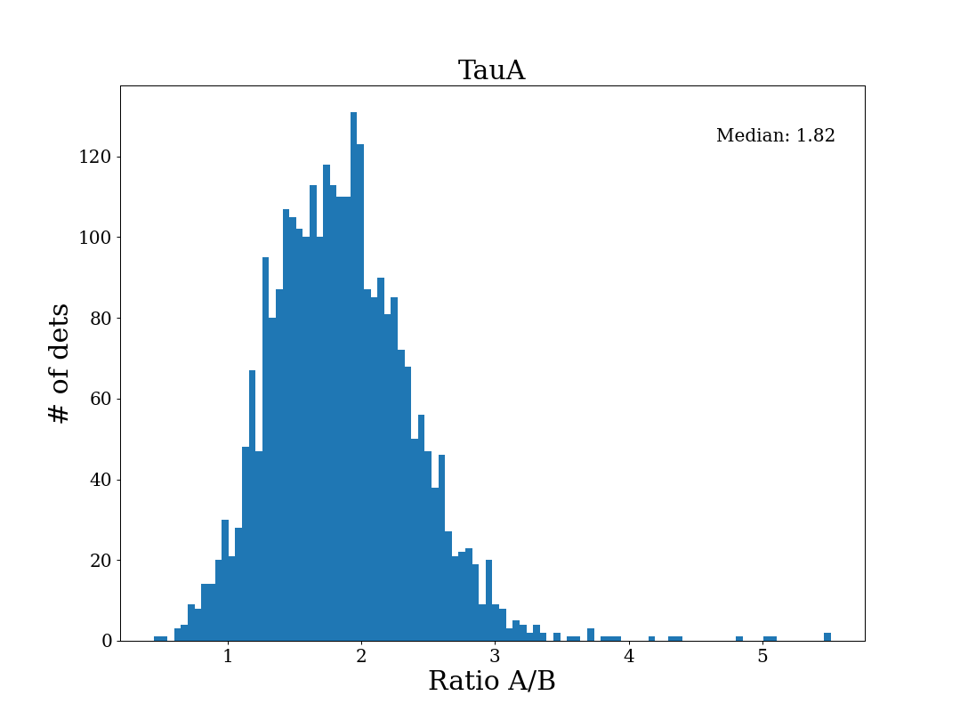}
    \caption{Left: Example $V_\textrm{coadd}$ map of a Tau~A observation showing noise estimate regions. We use pixels within eight circles around the center. They are defined to have a 0.1-degree radius, with centers at ($\pm0.25,0$), ($0,\pm0.25$), and ($\pm0.25,\pm0.25$) degree. The center region is used for the signal calculation. Right: Histogram of the ratio $\mathcal{N}_A/\mathcal{N}_B$ of two noise estimates for all detectors across all observations. The two noise estimates are defined in section \ref{subsec:map making}.  \edit{If the map pixels were uncorrelated, the two noise estimates would be expected to agree. The observed median ratio of about 1.8 reflects the pixel correlations along the scan direction and is used to correct $\mathcal{N}_B$.}}
    \label{fig:map noise}
\end{figure}

The statistical uncertainty is estimated from the map.
Figure~\ref{fig:map noise} (left) shows the regions from an example noise map used for noise estimation. Noise estimate regions are defined as having a 0.1-degree radius with the centers of ($\pm0.25,0$), ($0,\pm0.25$), and ($\pm0.25,\pm0.25$) degree. We perform two methods for noise calculation and compare the results.

\begin{itemize}
    \item Method A: RMS of eight regions\\
This method calculates the mean of each circle of the map first (figure~\ref{fig:map noise}) and then calculates the root mean square (RMS) of eight values. It is described as
\begin{equation}
    \mathcal{N}_A = \frac{1}{\sqrt{8}} \times \sqrt{\sum_{j=1}^8 \left( \frac{\sum_{i\in C_j} v_i}{N_{C_j}} \right)^2},
\end{equation}
where $v_i$ is the value of pixel $i$, $C_j$ is the circle $j$ ($j=1,2,...,8$), and $N_{C_j}$ is the number of pixels in $C_j$. This method directly accounts for the correlations among map pixels, but its statistical uncertainty is large because only eight regions are available.

\item Method B: RMS of all pixels in eight regions\\
This method calculates the RMS of all pixels in noise regions. It is described as
\begin{equation}
    \mathcal{N}_B = \frac{1}{\sqrt{N_{C_s}}} \times \sqrt{\frac{\sum_{i\in C} v_i^2}{N_C}},
\label{eq:map noise}
\end{equation}
where $v_i$ again is the value of pixel $i$, $N_C$ is the number of pixels of all noise calculated regions, and $N_{C_s}$ is the number of pixels in the signal calculated region. This method has a significantly smaller statistical error than method A, thanks to the approximately 700 pixels. However, the noise level of method B is underestimated due to pixel correlation in the azimuth direction, which is the telescope's scan direction. This can be seen in figure~\ref{fig:map noise}. Both blue and red pixels appear to be horizontally connected by about three pixels along the $x$ axis, whereas the $y$ axis is not, indicating a correlation in the azimuth direction.
\end{itemize}

As our final noise estimate, we calculate the ratio between the two methods for all detectors across all observations. We then take the median of them and multiply it by $\mathcal{N}_B$ as
\begin{equation}
    \mathcal{N} = \mathcal{N}_B \times \mathrm{median}(\mathcal{N}_A/\mathcal{N}_B).
    \label{eq:noise_of_v_frac_def}
\end{equation}
Figure~\ref{fig:map noise} (right) shows the histogram of the ratio of two methods of Tau~A observations.
The median value is about 1.8 for Tau~A, which is consistent with Jupiter and other sources reported in Paper~II. The value is reasonable considering the number of correlated pixels in figure~\ref{fig:map noise}. \edit{Here, $\mathcal{N}$ represents the statistical uncertainty in the mean $V_\textrm{coadd}$ signal within the source region. The statistical uncertainty in $V_\textrm{frac}$ can then be expressed as
\begin{equation}
    \sigma_\textrm{stat}
    =
    \frac{\mathcal{N}}{I_\textrm{source}} \:,
    \label{eq:sigma_stat}
\end{equation}
where the mean observed source intensity is defined as $I_\textrm{source} = \sum_p^{r<0.1} I_{\textrm{coadd},p}/N_{C_s}$.}


\section{Results}
\label{subsec:characterization}
The goal of this analysis is to determine calibration parameters and \edit{characterize the uncertainties relevant to circular polarization measurements}.
As demonstrated in figure~\ref{fig:optics schem demod}, our telescope optics with the addition of the QWP allow us to obtain data for intensity, for one linear polarization (Stokes $Q$) and for circular polarization (Stokes $V$).
We utilize both polarizations for validating our telescope and conducting systematics studies of astronomical sources.
Jupiter observations are used for I-to-V leakage validation, and Tau~A observations are used for linear polarization efficiency and U-to-V leakage estimations. I-to-V leakage uncertainty estimated from Jupiter observations is also validated with the Tau~A observations.
Note that Q-to-V leakage is negligibly small in this analysis, as described in section~\ref{subsubsec:circular polarization}.

\subsection{Jupiter observations}
\label{subsec:I-to-V uncertainty}
\begin{figure}[tbp]
    \centering
    \includegraphics[width=0.8\linewidth,keepaspectratio]{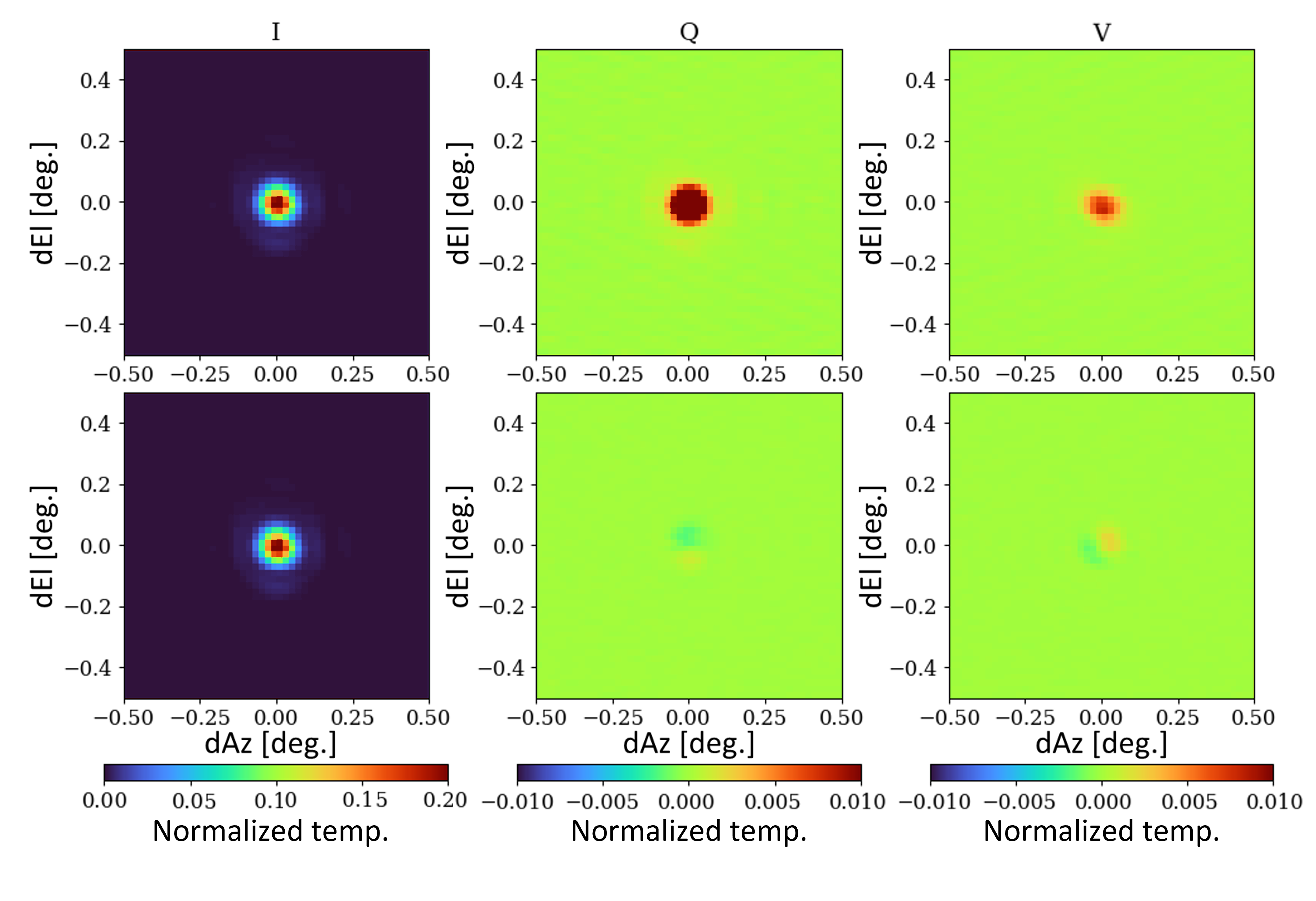}
    \caption{Jupiter maps with and without I-to-P leakage subtraction. Maps are coadded over the detectors of one observation. From left to right, observed intensity, linear polarization, and circular polarization are shown. The top row shows the map without the leakage subtraction, and the bottom row shows the map after subtraction. The color scale is normalized by Jupiter intensity. \edit{The comparison demonstrates that the leakage subtraction removes the dominant I-to-P leakage structure.}}
    \label{fig:Jupiter leakage coef}
\end{figure}

We use Jupiter observations to verify our overall pipeline and estimate the \edit{systematic uncertainty associated with the I-to-V leakage correction}. It is a bright, \edit{compact} point source widely used for telescope calibration, and its linear polarization is negligible for this work.
Figure~\ref{fig:Jupiter leakage coef} shows the maps of one Jupiter observation with and without I-to-P leakage subtractions.
We find that the leakage is successfully removed in the maps as a result of the data processing described in section~\ref{subsec:data processing}.
The typical value of the I-to-Q (I-to-V) leakage is about 6\% (1\%). \edit{These values are larger than those for} the PB-1 or PB-2a telescopes due to the QWP, but \edit{are} reasonable considering that the QWP has about \SI{2}{K} HWPSS~\cite{Takakura_2017}.

The relevant systematic associated with the I-to-V leakage is not the leakage itself, which is corrected observation by observation, but the uncertainty in that correction. Since the source intensity is much brighter than the typical circular polarization level, even a small misestimation of the leakage coefficient can produce a significant uncertainty in the final circular polarization measurement. The uncertainty in the I-to-V leakage correction \edit{has a statistical nature}, showing detector-to-detector and observation-to-observation variation rather than behaving as a fixed bias. This is consistent with its origin in detector non-linearity (section~\ref{subsec:data processing}). We therefore establish a model to estimate this uncertainty, assuming that Jupiter's circular polarization remains constant during our observation period. The same model is applied to Tau~A and other point sources, and its validity is supported by the Tau~A analysis presented below and by the null test results in Paper~II.

\edit{We model the uncertainty associated with the I-to-V leakage correction empirically using the Jupiter observations. As defined in section~\ref{subsec:map making}, $V_\textrm{frac}$ is the map-domain counterpart of $V_\textrm{obs}/I_\textrm{obs}$, and its statistical uncertainty $\sigma_\textrm{stat}$ is given by eq.~\eqref{eq:sigma_stat}. Since the low-frequency atmospheric intensity fluctuations used for the leakage estimate are approximately independent of the observed source, we assume that the uncertainty in the I-to-V leakage coefficient scales with $\mathcal{N} \, (=\sigma_\textrm{stat}\,I_\textrm{source})$, the statistical uncertainty in the source $V$ measurement. Thus, we model the I-to-V leakage-correction uncertainty in $V_\textrm{frac}$ as
\begin{equation}
    \sigma_\textrm{ItoV}
    =
    B\,
    \frac{I_\textrm{source}}{I_\textrm{Jupiter}}\,
    \sigma_\textrm{stat} \:,
    \label{eq:ItoV_uncertainty}
\end{equation}
where $B$ is a source-independent proportionality factor and $I_\textrm{Jupiter}$ is the mean observed Jupiter intensity defined in the same manner as $I_\textrm{source}$.
We also note that the I-to-V leakage-correction uncertainty is directly related to the uncertainty in the leakage coefficient as
$\sigma_\textrm{ItoV}=\sigma(\epsilon_{IV})$, since an error $\delta\epsilon_{IV}$ in the leakage coefficient produces a residual $\delta V_\textrm{obs}=\delta\epsilon_{IV}\,I_\textrm{obs}$, and hence $\delta(V_\textrm{obs}/I_\textrm{obs})=\delta\epsilon_{IV}$.}

\begin{figure}[tbp]
\centering
\includegraphics[height=6cm]{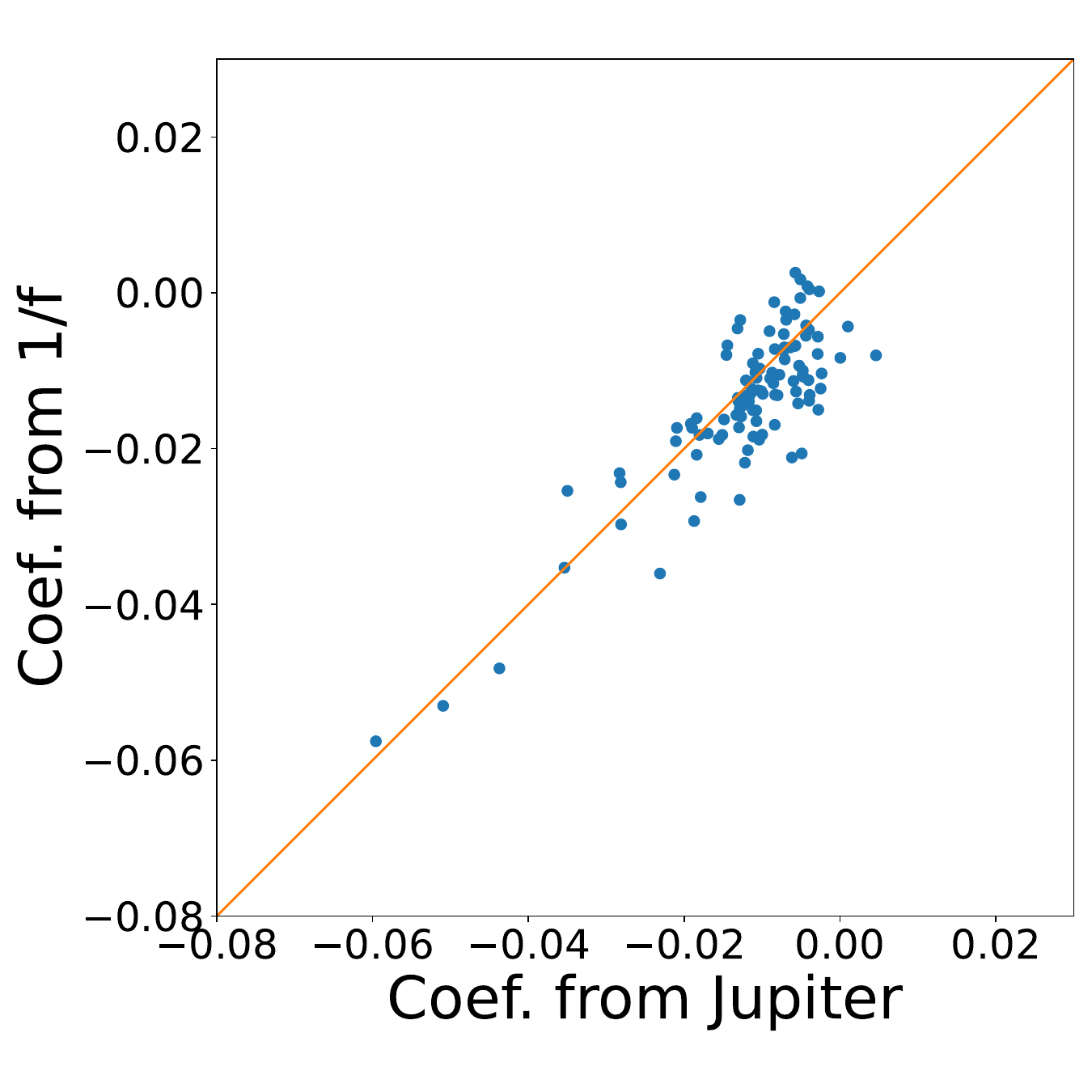}
\qquad
\includegraphics[width=.5\textwidth]{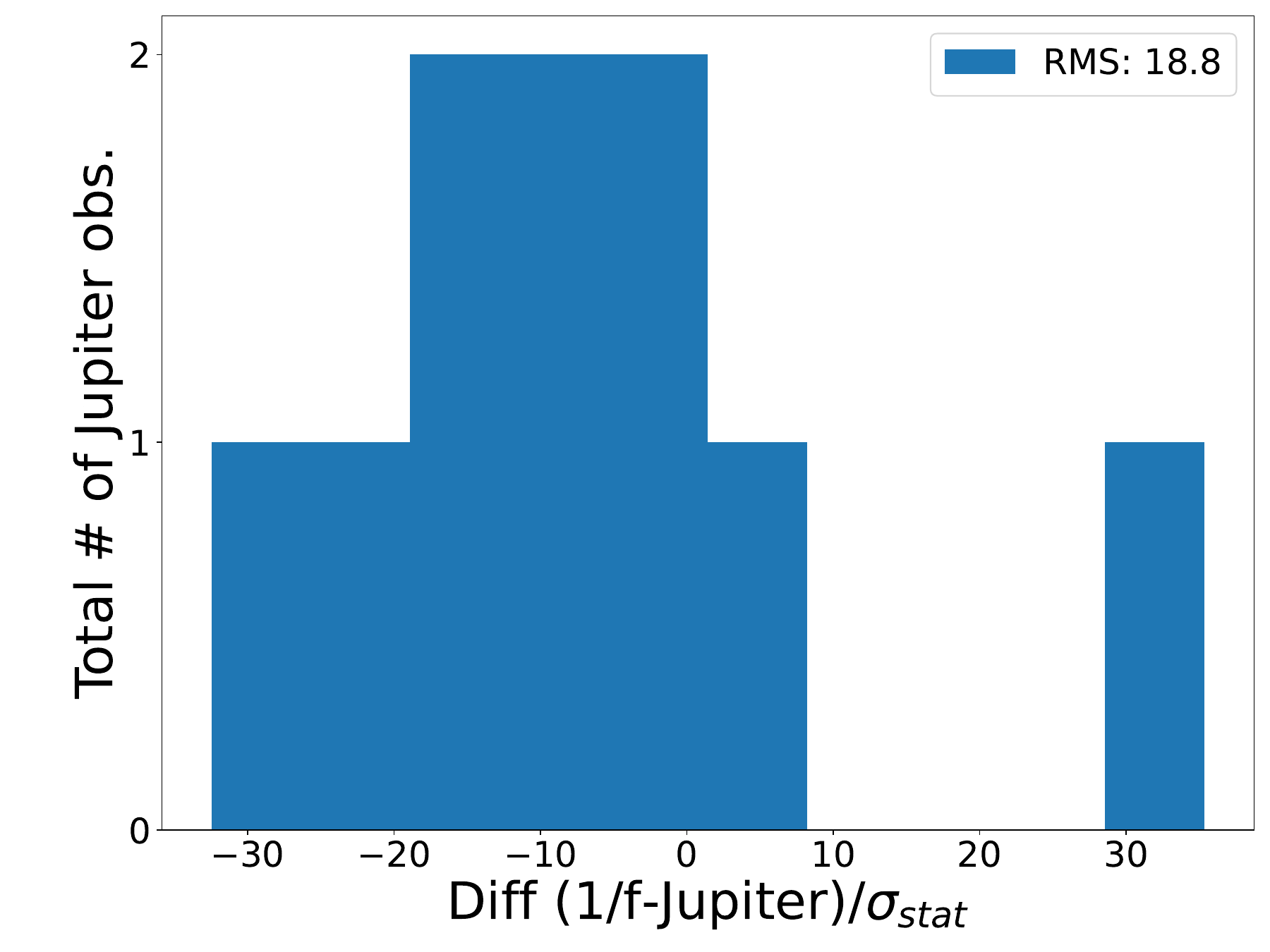}
\caption{Left: Comparison of the I-to-V leakage coefficients from the \edit{low-frequency component of TOD and Jupiter-map estimates} for one observation. The values obtained by the two different methods are in good agreement with each other. Right: \edit{Distribution of the differences between the two estimates normalized by $\sigma_{\textrm{stat},o}$ for multiple Jupiter observations. The RMS of this distribution gives $B=18.8$.}
\label{fig:Jupiter leakage}}
\end{figure}

\edit{The factor $B$ is determined using Jupiter observations by comparing two observation-level estimates of the I-to-V leakage coefficient~\cite{Yamada2024}. The first estimate is based on low-frequency fluctuations of the TOD and is used in the nominal data processing for the TOD-level leakage correction (section~\ref{subsec:data processing}); it is an estimate of the per-detector, per-observation leakage coefficient $\lambda_4^i$ in eq.~\eqref{eq:hwpss leakage}. For the comparison here, we calculate, for each observation $o$, the weighted mean of the estimated $\lambda_4^i$ over detectors and denote the resulting detector-averaged coefficient as $\bar{\lambda}_{4,o}^i$.  The second estimate is used only for Jupiter and serves here to characterize the uncertainty in the first estimate. It uses the Jupiter intensity and circular polarization maps. For each observation, we calculate the leakage coefficient from the detector-coadded Jupiter map by integrating the map within a 0.1-degree radius, and denote it as $\bar{\lambda}_{4_\textrm{Jupiter},o}^i$.\footnote{\edit{We use the {\it bar} to distinguish the detector-averaged quantities for each observation, $\bar{\lambda}_{4_\textrm{Jupiter},o}^i$ and $\bar{\lambda}_{4,o}^i$, from the per-detector, per-observation leakage coefficient, $\lambda_4^i$, for which the detector and observation indices are omitted.}} The left panel of figure~\ref{fig:Jupiter leakage} shows an example of the comparison for one observation, demonstrating good agreement between the two estimates.}

\edit{At the individual-observation level, the uncertainty represented by $\sigma(\epsilon_{IV})$ corresponds to the uncertainty in $\bar{\lambda}_{4,o}^i$.  Thus, applying $B=\sigma(\epsilon_{IV})/\sigma_\textrm{stat}$ to each observation, we estimate the factor $B$ as
\begin{equation}
    B
    =
    \sqrt{\frac{1}{N}\sum_o
    \left(
    \frac{\bar{\lambda}_{4,o}^i-\bar{\lambda}_{4_\textrm{Jupiter},o}^i}
         {\sigma_{\textrm{stat},o}}
    \right)^2} \:,
\end{equation}
where $\sigma_{\textrm{stat},o}$ is the statistical uncertainty in $V_\textrm{frac}$ defined in eq.~\eqref{eq:sigma_stat} for Jupiter observation $o$, and $N$ is the number of Jupiter observations. The right panel of figure~\ref{fig:Jupiter leakage} shows the distribution of the normalized differences entering this calculation, which yields $B=18.8$. The resulting uncertainty model in eq.~\eqref{eq:ItoV_uncertainty} is then applied to each source using its measured intensity relative to Jupiter. For Tau~A, the measured intensity ratio gives $B\,(I_{\textrm{obs,TauA}}/I_{\textrm{obs,Jupiter}})=2.8$.}

\subsection{Tau A observations}
\begin{figure}[tbp]
    \centering
    \includegraphics[width=0.9\linewidth,keepaspectratio]{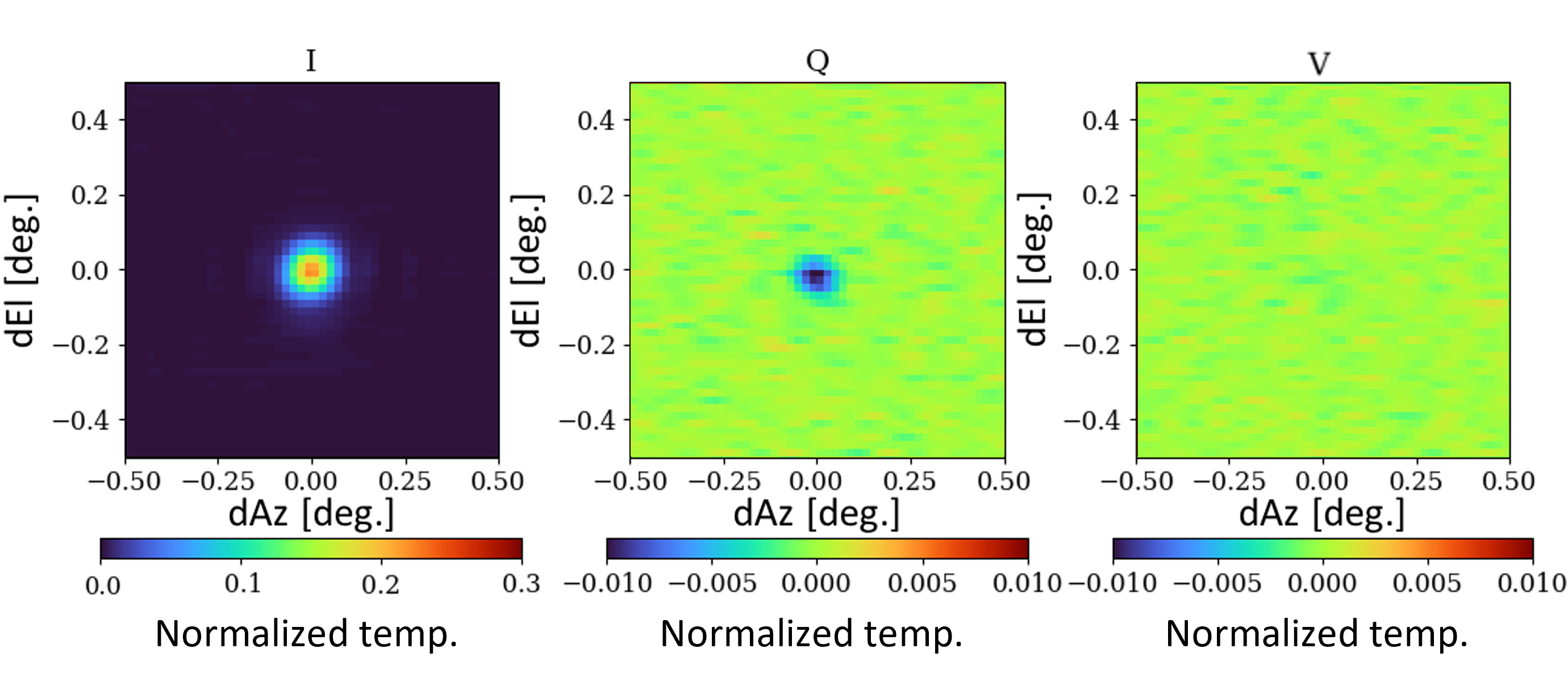}
    \caption{Tau A map of one observation. Maps are coadded over the detectors. From left to right, observed intensity, linear polarization, and circular polarization are shown. The map is after I-to-P leakage subtraction. The color \edit{scale} is normalized by Tau~A intensity.}
    \label{fig:TauA map}
\end{figure}

Tau~A observations are used to determine the linear polarization efficiency and U-to-V leakage of our system\edit{, and to provide a consistency check of the circular polarization measurement. In addition, the Tau~A analysis validates the I-to-V leakage uncertainty model, including its parameter determined from Jupiter observations in section~\ref{subsec:I-to-V uncertainty}.} Figure~\ref{fig:TauA map} shows an example of a coadded map of one Tau~A observation. The linear and circular polarization fractions are calculated for each observation.

\subsubsection{Efficiency calibration}
\begin{figure}[tbp]
\centering
\includegraphics[width=.47\textwidth]{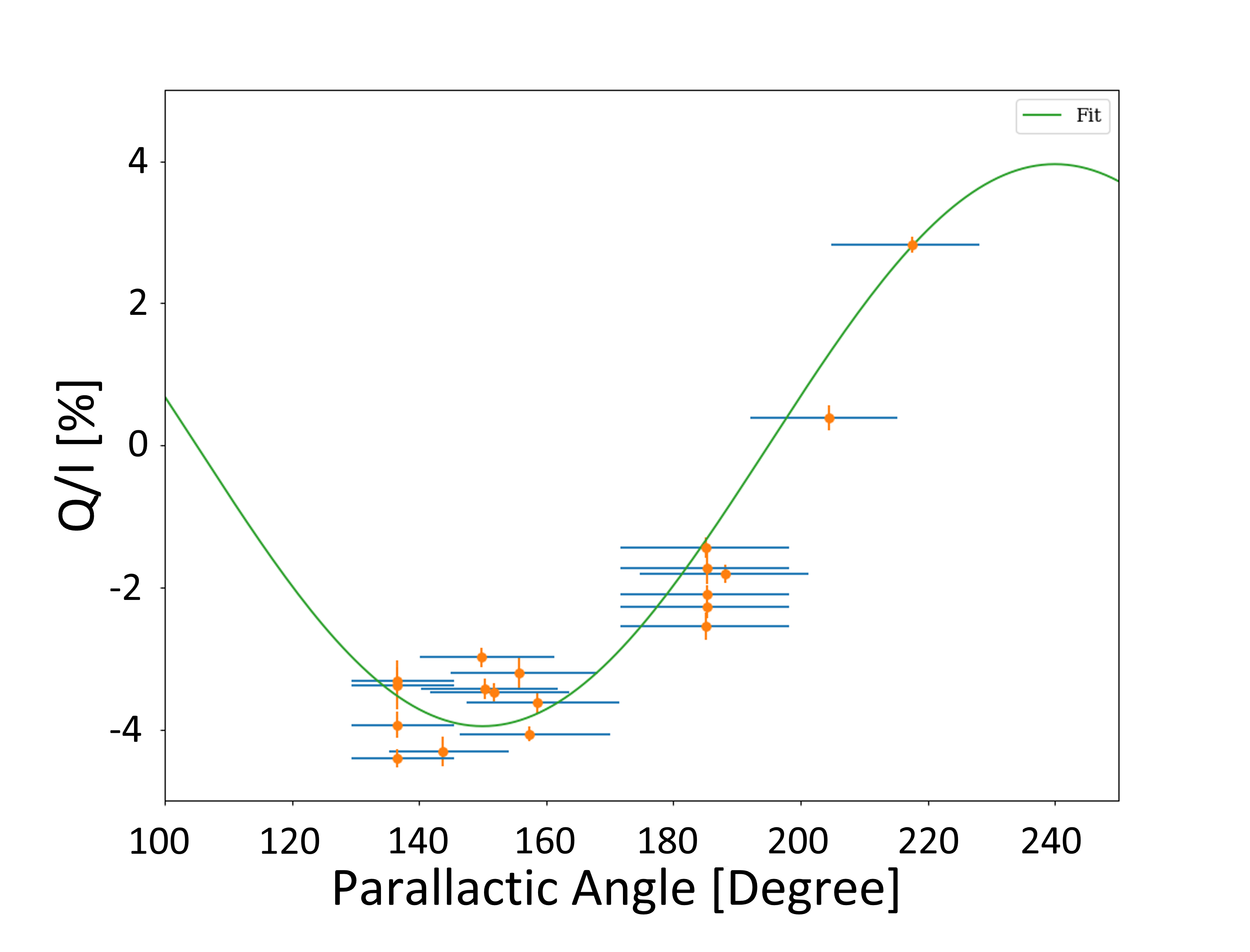}
\qquad
\includegraphics[width=.47\textwidth]{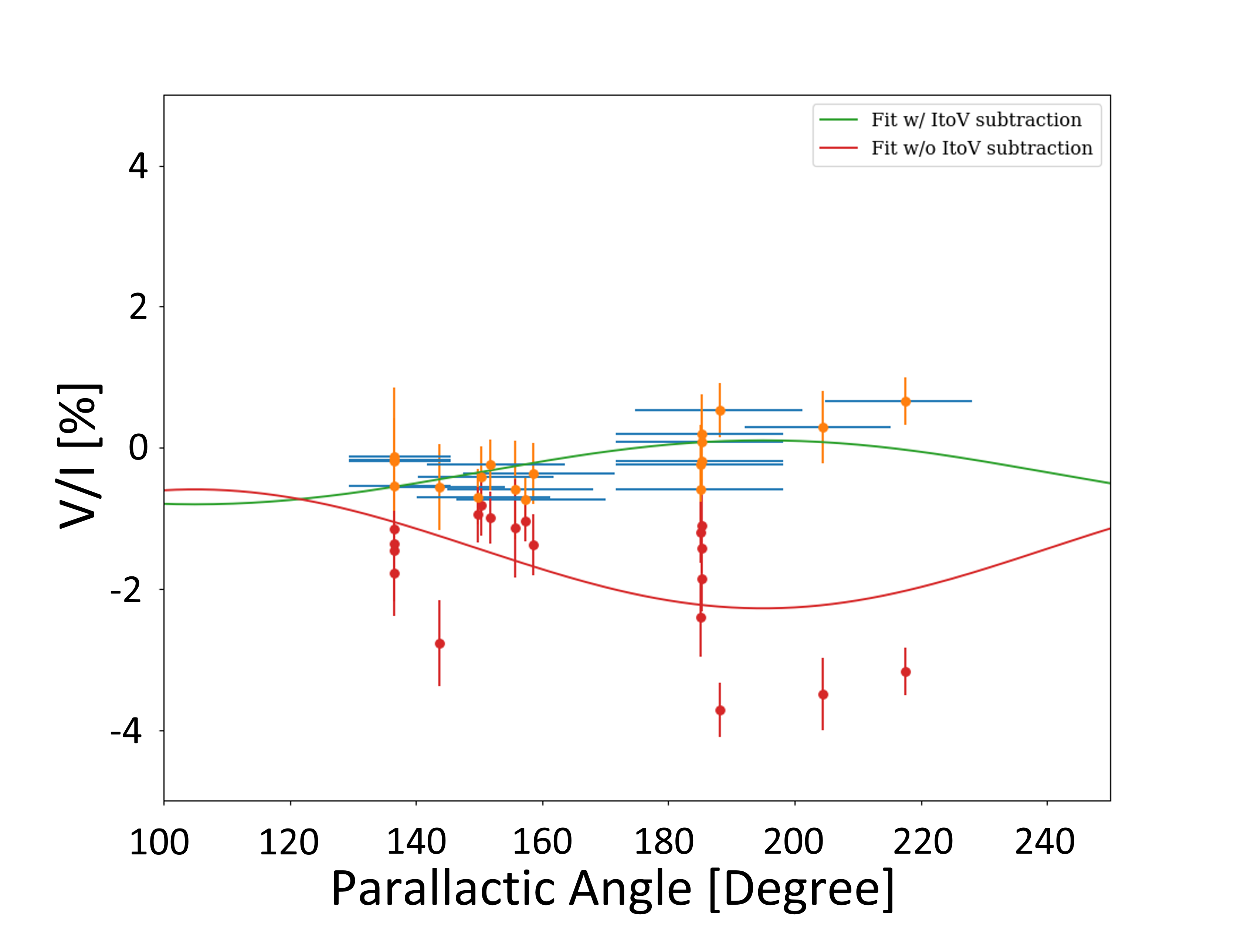}
\caption{Left: Tau~A linear polarization. The observed $Q$ signals in the telescope frames are shown. Each dot represents a ratio to the intensity of the coadd map of an observation. The parallactic angle is the mean angle when hitting Tau~A. The error bar is calculated from the coadd map with noise correction in. The blue line shows the parallactic angle during the observation, from start to end. The green line shows the fit curve. \edit{The sinusoidal dependence on parallactic angle is expected from the projection of the Tau~A linear polarization onto the telescope coordinates, and the fit yields a polarization efficiency of $46\pm2$\%.}  Right: Observed Tau~A circular polarization. Orange, green, and blue lines show \edit{the data and fit of the nominal analysis} in the same manner as linear polarization. \edit{In the circular-polarization fit, a sky Stokes $V$ signal appears as a parallactic-angle-independent offset, while U-to-V leakage produces the sinusoidal component. The nominal fit gives $V_{\textrm{frac}}=0.78\pm0.42$\% and $\epsilon_{UV}=-11.7\pm5.1$\%.} Red points and curve show the result without I-to-V leakage subtraction. \edit{The reduced chi-squared is 3.96 for the data without I-to-V leakage subtraction and 0.85 for the nominal analysis after the leakage subtraction and inclusion of the I-to-V leakage-correction uncertainty. The latter value indicates that the assigned uncertainty is consistent with the observed scatter.} \label{fig:TauA result}}
\end{figure}

We estimate the linear polarization efficiency, $\eta_Q$, \edit{using the Tau~A linear polarization measurement}. We assume that the Tau~A linear polarization fraction is 8.8\% and polarization angle is about $150^\circ$~\cite{TauA_linear_2010}.
The linear polarization measurement also serves to verify the polarization response relevant to the circular polarization search. As described in section~\ref{subsubsec:pol_opt_model}, the observed Stokes $Q$ and $V$ share the same polarization transfer function except for the QWP conversion factor.
Figure~\ref{fig:TauA result} (left) shows the linear polarization result of all Tau~A observations. The observed Stokes $Q$ shows a sinusoidal dependence on the parallactic angle across observations (figure~\ref{fig:optics schem demod}). Each data point represents one observation of Tau A, with the orange point indicating the ratio of linear polarization to intensity from the corresponding coadded map versus the mean parallactic angle for Tau A across the times it was being observed. We fit the observed $Q$ fraction ($Q_{\textrm{obs}}/I_{\textrm{obs}}$) as a function of parallactic angle using the efficiency as the only free parameter, and obtain an efficiency of $46\pm2$\%.

Several factors contribute to the inefficiency of the linear polarization measurement.
One contribution comes from optics, including the HWP and QWP. From the Mueller matrix calculation based on lab measurements, the efficiency of the HWP with AR coatings is $>99$\%. The QWP does not reduce the linear-polarization efficiency relative to its intensity transmission (eq.~\eqref{eq:mueller}). Thus, this combination suggests the efficiency $>99$\%.
Another instrument factor is the detector. Our receiver has lenslet-coupled sinuous antenna transition edge sensor bolometer arrays. The polarization efficiency of detectors is suggested to be about 95\%~\cite{Westbrook2018}.
Additionally, the detector time constant reduces the response. The efficiency is about 80\%\footnote{While we do not explicitly deconvolve the detector time constant in the TOD analysis, its dominant effects are effectively accounted for through calibration. The attenuation of the modulated signal amplitude is absorbed into the overall polarization efficiency calibration using Tau~A. The associated phase shift, which appears as an effective detector polarization angle rotation and could otherwise lead to leakage or calibration bias if it varies between observations, is mitigated through the observation-by-observation detector angle calibration using the HWPSS (section~\ref{subsec:characterization and calibration}).}, assuming the time constant $\tau$ of \SI{15}{ms}~\cite{Takakura_2017}. PB-2b time constant without the QWP is reported to be around \SI{12}{ms}~\cite{Farias2025}.
Inefficiency can also be introduced during offline analysis, especially by TOD filtering.
We apply filters in the reconstruction of intensity and polarization maps.
To estimate this effect, we simulate an ideal noise-free Tau~A signal using the actual Tau~A scan information.
The same TOD filters and mapmaking procedure are then applied as in the real data analysis.
The resulting polarization efficiency of the TOD filtering and mapmaking process is about 97\%.
Despite considering all the possible contributions discussed above, the observed inefficiency cannot be fully explained. As an additional check, we compared the detector physical angles with and without the QWP and obtained consistent results. \edit{A separate} Tau~A observation without the QWP \edit{yielded} an efficiency of about 50\%.  \edit{These checks further support the treatment in eq.~\eqref{eq:mvv_mqq_relation}, in which the measured Stokes-$Q$ efficiency is related to the Stokes-$V$ efficiency through the QWP conversion factor.  A complete accounting of the observed inefficiency} is left for future work.

The measured linear polarization efficiency can be related to that of circular polarization via $\eta_V = \eta_Q \, (\eta_{VU}/\eta_{QQ})$ according to the polarization optics model described in section~\ref{subsubsec:pol_opt_model}. As illustrated schematically in figure~\ref{fig:optics schem demod}, $\eta_{QQ}$ and $\eta_{VU}$ are determined by the QWP properties and correspond to $T$ and $\eta$ in eq.~\eqref{eq:mueller}, respectively. Their ratio, $\eta_{VU}/\eta_{QQ}=\eta/T$, was measured to be 99.0\% from the laboratory characterization of the QWP (section~\ref{subsec:design and lab measurements}). At the current level of precision, this correction is negligible. We therefore adopt the measured efficiency of $46\pm2$\% as the circular polarization efficiency.


\subsubsection{Leakage calibration}
\label{subsubsec:circular polarization}
Tau~A circular polarization measurement is utilized to characterize U-to-V leakages and to validate the uncertainty associated with the I-to-V leakage correction estimated from Jupiter observations. Tau~A circular polarization at \SI{90}{GHz} is reported to be $<0.2$\%~\cite{wiesemeyer2011}.
The uncertainty induced by the I-to-V leakage correction is further supported by the null test results for the other sources presented in Paper~II.

\edit{Figure~\ref{fig:TauA result} (right) shows the measured circular polarization fraction, $V_\textrm{frac}$, of the Tau~A observations as a function of parallactic angle, together with the fit. Dividing eq.~\eqref{eq:Vo_expression} by $I_\textrm{obs}=M_{II}I_\textrm{sky}$ gives
\begin{equation}
    \frac{V_\textrm{obs}}{I_\textrm{obs}}
    =
    \frac{M_{VV}}{M_{II}}
    \left[
        \frac{V_\textrm{sky}}{I_\textrm{sky}}
        +
        \epsilon_{UV}
        \frac{P_\textrm{sky}}{I_\textrm{sky}}
        \sin(2\alpha)
    \right],
    \label{eq:Vfrac_expression}
\end{equation}
where $M_{VV}/M_{II}=\eta\,\eta_P/T$ as described in section~\ref{subsubsec:pol_opt_model}. We fit the data with two free parameters: a constant offset and the amplitude of the sinusoidal component. The constant offset is converted to the Tau~A circular polarization fraction using the calibrated relative response $M_{VV}/M_{II}$. The sinusoidal amplitude is converted to the U-to-V leakage coefficient using the same relative response together with the known Tau~A linear polarization fraction. The fit yields a circular polarization fraction consistent with zero and a nonzero best-fit U-to-V leakage, as discussed below.}

\edit{We apply the I-to-V leakage-correction uncertainty model derived from Jupiter observations to the Tau~A fit. The uncertainty used in the fit after including the I-to-V leakage-correction uncertainty, denoted by $\sigma'$, is given by
\begin{equation}
    \sigma'
    =
    \sqrt{\sigma_\textrm{stat}^2+\sigma_\textrm{ItoV}^2}
    =
    \sigma_\textrm{stat}
    \sqrt{
        1+
        B^2
        \left(
        \frac{I_\textrm{source}}{I_\textrm{Jupiter}}
        \right)^2
    } \:.
\end{equation}
The resulting fit has a reduced chi-squared of 0.85, indicating that the assigned uncertainty is consistent with the observed scatter. For comparison, the fit to the data without I-to-V leakage subtraction has a reduced chi-squared of 3.96. This comparison supports both the I-to-V leakage correction and the associated uncertainty model derived from Jupiter observations.}  We use this \edit{uncertainty model in the other point-source analyses as well (Paper~II)}.

The fitted U-to-V leakage is $\epsilon_{UV}=-11.7\pm5.1$\%.
The value is consistent with that from the QWP lab measurement, which is $-7.2\%$ (see section~\ref{subsec:design and lab measurements}). \edit{Although the measured leakage is at the approximately 10\% level, its impact is reduced for sources with small linear polarization. The effect of U-to-V leakage can also be mitigated by the scan strategy (see Paper~II).}
The measured circular polarization fraction of Tau~A is $0.78\pm0.42$\%. This value does not indicate a significant detection of circular polarization in Tau~A, which is consistent with the previous publication~\cite{wiesemeyer2011}.

The Q-to-V leakage is expected to be much smaller than the U-to-V leakage. A detector angle miscalibration can rotate Stokes $Q$ into the detector-frame $U$ component. \edit{In the absence of such miscalibration, Stokes $V$ is converted by the QWP into the detector-frame $U$ component. Thus, the angle miscalibration appears as an effective Q-to-V leakage.} The Stokes $Q$ and $U$ in this context are defined in the coordinate system of the QWP birefringence axis. Since our detector polarization angle calibration uses the HWPSS, which primarily arises from the QWP differential emissivity aligned with this axis, this type of angle miscalibration is intrinsically suppressed. A small residual miscalibration can arise from the secondary HWPSS contribution due to polarization generated by the off-axis reflector telescope optics, but only when the telescope optical axis is misaligned with respect to the QWP birefringence axis. The measured residual QWP misalignment at the site is $<0.3^\circ$ (section~\ref{subsubsec:mounting}), corresponding to an effective Q-to-V leakage of less than 1.1\%, an order of magnitude smaller than the measured U-to-V leakage. We therefore neglect this effect in the analysis.

\section{Conclusion}
\label{sec:conclusion}
We have developed and characterized a method to search for circular polarization \edit{at millimeter wavelengths} by incorporating \edit{a QWP into the SA PB-2b optical system.  The QWP converts incident circular polarization into linear polarization before HWP modulation. This allows Stokes $V$ to be measured with the PB-2b's TES-bolometer-equipped receiver and its associated analysis pipeline, requiring only minor modifications to the existing infrastructure.
Laboratory measurements give a QWP conversion efficiency, $\eta/T$, of 99.0\% in the \SI{90}{GHz} band.}

Using Jupiter and Tau~A observations, we have established a calibration scheme and quantified \edit{the systematic uncertainties, including those associated with I-to-V and U-to-V leakage}. Jupiter observations are used to \edit{establish the I-to-V leakage-correction uncertainty model, including its scaling with source intensity relative to Jupiter}. \edit{The assigned uncertainty is supported by the Tau~A circular-polarization fit.} The linear polarization measurements of Tau~A confirm the overall consistency of our pipeline and provide an estimate of the effective polarization efficiency. The estimated \edit{polarization efficiency adopted for Stokes $V$ is $46\pm2$\%, and the measured U-to-V leakage is $\epsilon_{UV}=-11.7\pm5.1$\%. The Q-to-V leakage is estimated to be less than 1.1\% and is neglected in the analysis.}
The circular polarization results show no significant detection for Tau~A, \edit{with $V/I=0.78\pm0.42$\%,} which is consistent with the literature~\cite{wiesemeyer2011}. \edit{The calibration products established in this paper, including the Stokes-$V$ efficiency, I-to-V leakage-correction uncertainty prescription, U-to-V leakage characterization, and Q-to-V treatment, are applied to the circular-polarization search of other astronomical sources presented in Paper~II~\cite{mypaper}.}

\edit{In addition to the point-source observations, we also obtained CMB observation data with PB-2b for circular-polarization analysis. More broadly, the QWP-based method developed here may be applicable to other CMB polarization experiments, such as the Simons Observatory, that are primarily designed for linear-polarization measurements. The addition of a QWP provides a way to extend such systems to circular-polarization measurements with relatively minor modifications to the existing instrumentation and analysis framework.}

\acknowledgments

This work is supported by the Gordon and Betty Moore Foundation
Grant No.~GBMF7939, the Simons Foundation Grant No.~034079,
the John Templeton Foundation Grant No.~58724 and the National Science
Foundation grants AST-1440338 and AST-1207892.
This work was supported in part by the World Premier International
Research Center Initiative (WPI Initiative), MEXT, Japan.
In Japan, this work was supported by JSPS KAKENHI grant Nos.~JP18H05539,
JP19H00674, and JP23H00105, and the JSPS Core-to-Core Program JPJSCCA20200003.
Work at LBNL is supported in part by the U.S. Department of Energy,
Office of Science, Office of High Energy Physics, under contract
No.~DE-AC02-05CH11231.
Calculations were performed on the Central Computing System,
owned and operated by the Computing Research Center at KEK.
The Simons Array operated at the James Ax Observatory in the Parque
Astron\'omico Atacama in Northern Chile under the stewardship of the
Agencia Nacional de Investigaci\'on y Desarrollo (ANID).  Kana Sakaguri acknowledges support from the JSR Fellowship, the University of Tokyo;
the Forefront Physics and Mathematics Program to Drive Transformation (FoPM),
a World-leading Innovative Graduate Study (WINGS) Program, the University of Tokyo;
and the JSPS Research Fellowship for Young Scientists (DC),
JSPS KAKENHI Grant No.~JP24KJ0645.
Yuji Chinone acknowledges the support from JSPS KAKENHI Grant Number JP24K00667.  Masashi Hazumi thanks the Yushan Fellow Program by the Ministry of Education (MOE) of Taiwan.

The authors used OpenAI's ChatGPT (GPT-5 series models) during the preparation of this manuscript to assist with language editing and to improve the clarity and organization of the text. All AI-assisted output was reviewed and revised by the authors, who take full responsibility for the final content.



\bibliographystyle{JHEP}
\bibliography{biblio.bib}






\end{document}